\documentclass[12pt]{article}

\usepackage{amsmath} 
\usepackage{amsthm,amsmath,mathrsfs,amsfonts,amssymb}
\usepackage{graphicx,psfrag,epsf,fixmath,caption,subcaption}
\usepackage{enumerate}
\usepackage{natbib}
\usepackage{url}

\usepackage{enumitem} 

\newcommand{\blind}{1}
\newtheorem{theorem}{Theorem}

\theoremstyle{definition}

\newtheorem{InnerExample}{Example}
\newenvironment{example}
  {\InnerExample} 
  {\hfill $\square$ \endInnerExample}

\newtheorem*{InnerContExample}{Example \continuanceref~(cont'd)}
\newenvironment{ContExample}[1]
  {\newcommand\continuanceref{#1}\InnerContExample} 
  {\hfill $\square$ \endInnerContExample}

\newcommand{\cartproduct}{\times}
\newcommand{\pinv}{\star}

\DeclareMathAlphabet{\mathbbmsl}{U}{bbm}{m}{sl}

\begin{document}

\def\spacingset#1{\renewcommand{\baselinestretch}{#1}\small\normalsize} \spacingset{1}

\if1\blind
{
\title{\bf {\large FIXED-EFFECT REGRESSIONS ON NETWORK DATA}}
\author{ Koen Jochmans\thanks{Address: University of Cambridge, Faculty of Economics, Austin Robinson Building, Sidgwick Avenue, Cambridge CB3~9DD, U.K. E-mail: \texttt{kj345@cam.ac.uk}.}\\ University of Cambridge
 \and \setcounter{footnote}{2}
 Martin Weidner\thanks{Address: University College London, Department of Economics,  Gower Street, London WC1E~6BT, U.K., and CeMMAP. E-mail: \texttt{m.weidner@ucl.ac.uk}. 
\smallskip
\newline We are grateful to Ulrich M\"uller and three referees for constructive comments.  We would also like to thank Nadine Geiger, Bryan Graham, \'{A}ureo de Paula, Fabien Postel-Vinay, Valentin Verdier, and, in particular, Jean-Marc Robin for discussion and for comments on earlier drafts of this paper. Valentin Verdier also provided indispensable support that enabled us to include our illustration on the estimation of teacher value-added.
\newline 
Jochmans gratefully acknowledges financial support from the European Research Council through Starting Grant n\textsuperscript{o} 715787. Weidner gratefully acknowledges support from the Economic and Social Research Council through the ESRC Centre for Microdata Methods and Practice grant RES-589-28-0001 and from the European Research Council grant ERC-2014-CoG-646917-ROMIA. 
\newline
The first version of this paper dates from August 4, 2016. All versions are available at {\rm https://arxiv.org/abs/1608.01532}.
} \\  University College London
    }
  \maketitle
} \fi

\if0\blind
{
  \bigskip
  \bigskip
  \bigskip
  \begin{center}
    {\bf {\Large FIXED-EFFECT REGRESSIONS ON NETWORK DATA}}
\end{center}
  \medskip
} \fi

\medskip
\vspace{-.75cm}

\begin{abstract}
\noindent
This paper considers inference on fixed effects in a linear regression model estimated from network data. An important special case of our setup is the two-way regression model. This is a workhorse technique in the analysis of matched data sets, such as employer-employee or student-teacher panel data. We formalize how the structure of the network affects the accuracy with which the fixed effects can be estimated. This allows us to derive sufficient conditions on the network for consistent estimation and asymptotically-valid inference to be possible. Estimation of moments is also considered. We allow for general networks and our setup covers both the dense and sparse case. We provide numerical results for the estimation of teacher value-added models and regressions with occupational dummies.



\end{abstract}

\newpage

\spacingset{1.45}

\renewcommand{\thetable}{\Roman{table}} 

\setcounter{equation}{0}

\section{Introduction}

Data on the interaction between agents are in increasing supply. A workhorse technique to analyze such data is a linear regression model with agent-specific parameters. 
It has been used to investigate a variety of questions. 
For example, application of a two-way regression model to matched employer-employee data decomposes (log) wages into worker heterogeneity, firm heterogeneity, and residual variation. Following \cite{AbowdKramarzMargolis1999} the correlation between the estimated worker and firm effects is regarded as a measure of assortative matching. A positive correlation indicates that high quality workers are employed in more productive firms. Using the same decomposition, \cite{CardHeiningKline2013} study to what extent the evolution of wage inequality is due to changes in the variance of worker and firm heterogeneity. \cite{Nimczik2018} reports the whole distribution of the estimated worker and firm effects. In a similar fashion, the literature on student achievement backs out student and teacher effects from test score data. The estimated teacher heterogeneity is interpreted as teacher value-added and their variance as a measure of their importance (see \citealt{JacksonRockoffStaiger2014} for an overview of this literature). These estimates are used to assess teachers and are important inputs to personnel evaluations and merit pay programs \citep{Rothstein2010}.\footnote{Fixed-effect regressions of this kind are now part of the standard toolkit of many empiricists in a variety of different areas. 
\cite{FinkelsteinGentzkowWilliams2016} and \cite{AmitiWeinstein2018} use them to separate supply and demand factors in healthcare utilization from data on patient migration, and in firm investment behavior from financial data on banks loans, respectively. \cite{ChettyHendren2018} evaluate the importance of growing up in a specific neighborhood on labor market outcomes later on in life.}

In spite of their widespread use, there is little to no work on the theoretical properties of such fixed-effect approaches. In fact, the few results that are available point to issues of downward bias in the estimation of the correlation between worker and firm effects, finding a spurious negative correlation in many data sets  \citep{AndrewsGillSchankUpward2008,AndrewsGillSchankUpward2012}, and upward bias in the estimator of the variance of teacher effects \citep{Rockoff2004}. The presence of bias here is not surprising. Indeed, the individual effects are estimated with noise. Their sampling error then introduces bias in the estimator of nonlinear functionals. A more complicated issue is the assessment of the statistical precision with which the fixed effects are estimated and, more generally, the development of distribution theory. This is important as it allows to establish conditions for consistency and rates of convergence, and yields insight into whether standard test statistics can be expected to be approximately size correct and have non-trivial power. None of these issues has been addressed so far. Providing such theory so is not only relevant for inference on the individual effects and their moments, but may also serve as a stepping stone to address related problems. For example, without theory for the fixed-effect estimator the behavior of the falsification test for value-added models of \cite{Rothstein2010} remains unknown,
and correct standard errors for regressions of outcomes on estimated fixed effects \citep{KettemannMuellerZweimuller2017} cannot be derived.

The data structure arising from interactions between agents is different from that of standard cross-section or panel data. It is typically difficult to see how the data carry information about certain parameters. In this paper we present sufficient conditions for consistency and asymptotic normality of least-squares estimators of fixed effects in linear regression models. We see the data as a network and represent it by a graph where agents are vertices and edges between vertices are present if these agents interact. It is intuitive that the structure of this graph should be a key determinant of the accuracy of statistical inference. We formalize this here. 
Our setup places no a priori restrictions on the graph structure and our results apply to both dense and sparse settings. 
A data structure of particular importance is that of a bipartite graph. Here, the data concerns two types of individuals and interactions only occur between the types but not within each type. This is the case in our motivating examples above and we treat this bipartite case in detail.
In fact, while we deal with general graphs, our regression setup is designed to capture the main features of the prototypical two-way regression model. We focus on inference on the individual effects but our results also serve as a stepping stone for the analysis of estimators of other parameters, such as the variance and other moments of (the distribution of) the individual effects, 
and we provide some results on these as well.  We do not discuss inference on common slope coefficients. In contemporary work, \cite{Verdier2016} provides such results for two-way regression models. The fixed-effect model for test scores, for example,  can be used to assess the effect of class-size reductions on student achievement while controlling for student and teacher heterogeneity.

The ability to accurately estimate the individual effect of a given vertex depends on how well this vertex is connected to the rest of the network. Our theory involves both global and local measures of network connectivity. The main global connectivity measure we use is the smallest non-zero eigenvalue of the (normalized) Laplacian matrix of the graph.\footnote{The Laplacian matrix is similar to the adjacency matrix as a devise to represent a graph and can be obtained from it. Both matrices are formally defined below. 
Eigenvalues and eigenvectors of these and related matrices have also been found of use in determining equilibrium conditions in games played on networks \citep{BramoulleKrantonDamours2014} and in (statistical) community detection \citep{SchiebingerWainwrightYu2015}.} It reflects how easy it is to disconnect a network by removing edges from it. The other measures of connectivity that we use are the degrees of the vertices as well as various harmonic means thereof. All of these measures arise naturally when studying the variance of the fixed-effect estimator. We highlight the interplay between them in deriving conditions for consistent estimation and for inference based on standard first-order asymptotics to be possible. As the network grows the smallest eigenvalue may approach zero, and so the graph may become more sparse, provided the relevant harmonic mean grows sufficiently fast. These findings mimic conditions on the bandwidth in nonparametric estimation problems, although they will typically show up in second-order terms here.
This explains why estimation at the parametric rate may be feasible even in sparse networks. 
Our results also show that inference on averages over the individual effects is more demanding on the network structure and, even after bias reduction, may only be feasible in quite dense networks.

Our analysis shows that it is useful to inspect measures of global and local connectivity when interpreting estimation results from network data. We do so here for two data sets. The first is a large network of teachers in elementary schools in North Carolina, where the object of interest would be teacher value-added. This is arguably one of the most important applications of the two-way regression model. This graph is only very weakly connected and our theory does not support the use of large-sample arguments. When a simple model with homoskedastic errors is applied to these data 
standard errors based on conventional first-order approximations
for teacher value-added are, on average, about $40\%$ smaller than the actual standard deviations. Further, the sample variance of the estimated teacher effects has a substantial upward bias. This bias translates into an overly optimistic view on the ability of teacher value-added to explain variation in test scores. 
To provide an example of a data set that yields a much stronger connected graph we also construct an occupational network from the British Household Panel Survey (BHPS). This graph would arise in the context of wage regressions with occupational dummies, for example. Here, our connectivity measures are much more supportive of standard inferential approaches and, indeed, again in a simple model, we find that conventional first-order approximationa are quite accurate.

The remainder of this paper is organized as follows. Section 2 details the structure of the data under study and introduces the regression model of interest. Special attention is given to the bipartite graph and the two-way regression model. Section 3 provides distribution theory for the least-squares estimator of the individual effects and also discusses estimation of their moments. Section 4 contains details on our two numerical illustrations. Section 5 concludes. The supplementary material to this paper contains some additional results and illustrations, as well as the proofs of all theorems.

\section{Regression analysis of network data} 

\subsection{Data structure}

Consider an undirected graph $\mathcal{G}:=\mathcal{G}(V,E)$ where $m := \lvert E \rvert$ edges are placed between $n := \lvert V \rvert$ vertices. We allow for multiple edges between vertices (i.e., $\mathcal{G}$ can be a multigraph) and the edges may be assigned a weight. We do not consider loops (i.e., no edge connects a vertex with itself). Without loss of generality we label the vertices by natural numbers, so that $V$ is $\{1,\ldots, n\}$.
The multiset $E$ contains the $m>0$ unordered pairs $(i,j)$ from the product set $V\cartproduct V$ that are linked by an edge, possibly with repetition. The same pair $(i,j)$ will appear multiple times in $E$ if they share more than one edge; we let ${E}_{(i,j)}\subset E$ denote the set of edges between them. We have $E_{(i,j)}=E_{(j,i)}$ and may have $E_{(i,j)}=\varnothing$. We will label the edges by natural numbers; so, each edge $e$ edge has assigned to it an integer $\varepsilon_e\in \lbrace 1,\ldots,m \rbrace$. 
For later use we note that vertices $i$ and $j$ are said to be connected if $\mathcal{G}$ contains a path from $i$ to $j$, and that the graph $\mathcal{G}$ is said to be connected if every pair of vertices in the graph is connected.

For an edge $e\in E$ let $w_e>0$ be its weight. An unweighted graph has $w_e=1$ for all $e\in E$. The graph $\mathcal{G}$ may be represented by its $m\times n$ (oriented) incidence matrix $\mathbold{B}$, with entries
\begin{equation} \label{eq:incidence}
(\mathbold{B})_{\varepsilon_e i} := 
\left\lbrace \begin{array}{rl}
\sqrt{w_e} & \text{if $e\in E_{(i,j)}$ for some $j\in V$ and $i<j$,}
\\
-\sqrt{w_e} & \text{if $e\in E_{(i,j)}$ for some $j\in V$ and $i>j$,}
\\
 0 & \text{otherwise.} 
\end{array}\right. 
\end{equation}
Here, the choice of sign gives each edge $e$ an orientation. As will become apparent, our analysis below is invariant to this choice
of orientation.
The graph may also be represented through its $n\times n$ adjacency matrix $\mathbold{A}$, which has elements
$$
(\mathbold{A})_{ij}
:= 
\sum_{e\in E_{(i,j)}} w_e  .
$$
The incidence matrix and adjacency matrix are related through the $n\times n$ Laplacian matrix $\mathbold{L}$, as 
$$
\mathbold{L}:= \mathbold{B}^\prime\mathbold{B} = \mathbold{D}-\mathbold{A},
$$
for $\mathbold{D}:=\mathrm{diag}(d_1,\ldots,d_n)$ the diagonal $n\times n$ (weighted) degree matrix, where the degree of vertex $i$ is 
$$
d_i :  = \sum_{j=1}^n (\mathbold{A})_{ij}.
$$
When $\mathcal{G}$ is an unweighted graph, for example, $d_i$ equals the number of edges that involve vertex $i$. For a vertex $i$ we will let 
$
[i] := \lbrace j\in V: E_{(i,j)}\neq \varnothing \rbrace
$
denote the set of its direct neighbors. Observe that $d_i$ may be large even if $i$ has few neighbors---i.e, when $\lvert [i] \rvert$ is small---as the edge weights $(\mathbold{A})_{ij}$ for $j\in [i]$ may be large. An example is a multigraph where many edges exist between $i$ and some $j\in [i]$.

\subsection{Regression model and least-squares estimator}

Now, given a graph $\mathcal{G}$, for each edge $e\in E$ we observe an outcome $y_{\varepsilon_e}$ and a $p$-vector of covariates $\mathbold{x}_{\varepsilon_e}$. Allowing $\mathcal{G}$ to be a multigraph covers the (unbalanced) panel data case, where multiple outcomes are available for some vertex pairs. Collect all outcomes in the $m$-vector $\mathbold{y}$ and all covariates in the $m\times p$ matrix $\mathbold{X}$. Let $\mathbold{\alpha}:=(\alpha_1,\ldots,\alpha_n)^\prime$ be an $n$-vector of vertex-specific parameters and let $\mathbold{\beta}:=(\beta_1,\ldots,\beta_p)^\prime$ be a $p$-vector of regression slopes. Our interest lies in estimating the model
\begin{equation} \label{eq:regression1}
\mathbold{y} = \mathbold{B}\mathbold{\alpha} + \mathbold{X}\mathbold{\beta} + \mathbold{u},
\end{equation}
where $\mathbold{u}$ is an $m$-vector of regression errors.\footnote{A change of edge orientation corresponds to a sign flip in the corresponding outcome and regressor matrices. This  does not affect the least squares estimator of \eqref{eq:regression1}.} We will treat $\mathbold{B}$ and $\mathbold{X}$ as fixed throughout.
This implies that we consider the network as non-random
and exogenous.\footnote{Exogeneity of the network is the standard assumption in the literature building on \cite{AbowdKramarzMargolis1999}. 
Accounting for endogenous network formation requires more complicated models and has started to receive some attention; see \cite{BonhommeLamadonManresa2015} and \cite{LentzPiyapromdeeRobin2018}.
} 
Our focus is on the vector $\mathbold{\alpha}$. In the two-way regression model of our motivating examples these are the worker and firm effects or the student and teacher effects, respectively. By the definition of $\mathbold{B}$ one of these effects will enter \eqref{eq:regression1} with a minus sign. While this may appear to be an unusual convention from an applied perspective, it is convenient for our theoretical analysis. As will be explained below, this sign convention is without loss of generality in the two-way model, where the underlying graph is bipartite.

In \eqref{eq:regression1} the outcomes for a given vertex pair $(i,j)$ depend on the individual effects through their difference $\alpha_i - \alpha_j$. This implies that our model is overparameterized. Indeed, we have $\mathbold{B} \mathbold{\iota}_n =0$,
where $\mathbold{\iota}_n:=(1,\ldots,1)^\prime$ is the $n$-vector of ones, as each row of $\mathbold{B}$ sums up to zero. It follows that the mean of the vertex-specific parameters cannot be learned from the data and a normalization is required. We impose that 
\begin{equation} \label{eq:constraint}
\sum_{i=1}^n \sum_{j=1}^n (\mathbold{A})_{ij}\, (\alpha_i + \alpha_j) 
= 0,
\end{equation}
which will prove a convenient choice for our purposes. Denoting the degree vector by $\mathbold{d}:=(d_1,\ldots, d_n)^\prime$ we may write the constraint \eqref{eq:constraint} compactly as $\mathbold{d}^\prime \mathbold{\alpha}=0$. 
A normalization can be dispensed with if interest lies in parameter differences, i.e, $\alpha_i-\alpha_j$, as in \cite{FinkelsteinGentzkowWilliams2016}, for example. Results for such differences that parallel those developed below are given in the supplementary material.

The standard estimator of $\mathbold{\alpha}$ is the constrained least-squares estimator
\begin{equation} \label{eq:ols}
\check{\mathbold{\alpha}}
:=  
(\check{\alpha}_1,\ldots,\check{\alpha}_n)^\prime = 
\; \;
\arg\hspace{-.9cm}
\min_{\mathbold{a}\in \lbrace \mathbold{a}\in\mathbb{R}^n: \,
 \mathbold{d}^\prime \mathbold{a}  = 0
 \rbrace}
\;\;
\left\lVert \mathbold{M}_{\mathbold{X}}\mathbold{y}-\mathbold{M}_{\mathbold{X}}\mathbold{B}\mathbold{a} \right\rVert^2,
\end{equation}
where $\lVert\cdot \rVert$ denotes the Euclidean norm, 
$
\mathbold{M}_{\mathbold{X}}:= \mathbold{I}_m - \mathbold{X}(\mathbold{X}^\prime \mathbold{X})^{-1} \mathbold{X}^\prime,
$
and $\mathbold{I}_m$ is the identity matrix of dimension $m \times m$ .
The following theorem gives conditions under which this estimator exists and is unique. For any matrix $\mathbold{C}$ we denote its  Moore-Penrose pseudoinverse by $\mathbold{C}^\dagger$. When $\mathbold{C}$ is $n\times n$ we let $\mathbold{C}^\pinv := \mathbold{D}^{-1/2} \left( \mathbold{D}^{-1/2}  \mathbold{C}  \mathbold{D}^{-1/2}\right)^\dagger \mathbold{D}^{-1/2}$.
It is easily shown that $\mathbold{C} \mathbold{C}^\pinv \mathbold{C} = \mathbold{C}$
 and $  \mathbold{C}^\pinv \mathbold{C} \mathbold{C}^\pinv = \mathbold{C}^\pinv $. Therefore, 
 $\mathbold{C}^\pinv$ is a pseudoinverse of  $\mathbold{C}$.

\begin{theorem}[Existence]
\label{theorem:existence}
Let $\mathcal{G}$ be connected, $\mathrm{rank}(\mathbold{X})=p$, and $\mathrm{rank}((\mathbold{X},\mathbold{B}))=p+n-1$. Then
$$
\check{\mathbold{\alpha}} = (\mathbold{B}^\prime\mathbold{M}_{\mathbold{X}}\mathbold{B})^{\pinv} \mathbold{B}^\prime \mathbold{M}_{\mathbold{X}}\mathbold{y} 
$$
and is unique.
\end{theorem}

\noindent
The need for a pseudoinverse arises because $\mathbold{B}^\prime \mathbold{M}_{\mathbold{X}} \mathbold{B}$ is singular, which follows from the fact that $\mathbold{B} \mathbold{\iota}_n =0$. The use of the particular pseudoinverse  $(\mathbold{B}^\prime\mathbold{M}_{\mathbold{X}}\mathbold{B})^{\pinv}$ is a consequence of our normalization 
 $\mathbold{d}^\prime \mathbold{\alpha} = 0$. A change of normalization would imply a
different pseudoinverse in the statement of Theorem~\ref{theorem:existence}.
 The result of the theorem is intuitive and generalizes results in the literature on matched employer-employee data (\citealt{AbowdCreecyKramarz2002}). 
When the graph $\mathcal{G}$ is disconnected a separate normalization of the form in \eqref{eq:constraint} is needed for each 
connected component of $\mathcal{G}$. 
Our results then apply to each of these subgraphs. In practice, the analysis is typically confined to the largest connected component of $\mathcal{G}$ (see, for example, \citealt[p.~988]{CardHeiningKline2013}).

While $\check{\mathbold{\alpha}}$ is routinely used its statistical properties are not well understood. Our aim here is to shed light on how the structure of the network $\mathcal{G}$ affects its sampling behavior and, with it, the reliability of standard inferential procedures based on $\check{\mathbold{\alpha}}$.
For our analysis edge-specific covariates mostly complicate notation and presentation. It will on occasion be convenient to first analyze \eqref{eq:regression1} when $\mathbold{\beta}$ is treated as known and the outcome vector is redefined as $\mathbold{y}-\mathbold{X}\mathbold{\beta}$.  Then
$$
{\mathbold{\hat{\alpha}}} := (\mathbold{B}^\prime\mathbold{B})^{\pinv} \mathbold{B}^\prime (\mathbold{y}-\mathbold{X}\mathbold{\beta} )
$$
is the least-squares estimator of $\mathbold{\alpha}$ subject to \eqref{eq:constraint}. To appreciate how the structure of $\mathcal{G}$ relates to our problem of estimating the parameter $\mathbold{\alpha}$ suppose first that $\mathbold{u}\sim(\boldsymbol{0},\sigma^2 \mathbold{I}_m)$.
Then
\begin{equation} \label{eq:var0}
\mathrm{var}(\hat{\mathbold{\alpha}})
=
\sigma^2 \mathbold{L}^*.
\end{equation}
So, up to a scale factor, the variance of $\hat{\mathbold{\alpha}}$ is completely determined by the Laplacian of $\mathcal{G}$. If, in addition, we were to assume that $\mathbold{u}\sim N(\boldsymbol{0},\sigma^2\mathbold{I}_m)$ we would be in the classical regression setting and, given unbiasedness of $\hat{\mathbold{\alpha}}$, size-correct inference could be performed for any sample size. It is not clear, however, how one should proceed with non-classical regression errors.

The validity of standard large-sample arguments is not immediate here. From \eqref{eq:var0} we have
\begin{equation} \label{eq:var1}
{\rm var}( \hat{\alpha}_i )
= \sigma^2 \frac{(\mathbold{S}^\dagger)_{ii} }{d_i} ,
\end{equation}
where 
$$
\mathbold{S}:=\mathbold{D}^{-1/2} \mathbold{L} \mathbold{D}^{-1/2}
=
\mathbold{I}_n - \mathbold{D}^{-1/2} \mathbold{A} \mathbold{D}^{-1/2}
$$
is the normalized Laplacian. Equation \eqref{eq:var1} follows from the fact that $\mathbold{L}^*=\mathbold{D}^{-1/2}\mathbold{S}^\dagger\mathbold{D}^{-1/2}$, such that $(\mathbold{L}^\pinv)_{ii} = (\mathbold{S}^\dagger)_{ii}/d_i$.\footnote{Our choice of normalization \eqref{eq:constraint} guarantees the appearance of the Moore-Penrose pseudoinverse of $\mathbold{S}$ in $\mathbold{L}^*$, which is the main reason for that choice.
We are grateful to Nadine Geiger for pointing out an inconsistency in our normalization in an earlier version of this paper.
}
While \eqref{eq:var1} shows the importance of the sample size in the variance of $\hat{\alpha}_i$ through the presence of the degree $d_i$, it does not imply that ${\rm var}( \hat{\alpha}_i )$ shrinks as $d_i\rightarrow \infty$, nor would it give a convergence rate if it did. This is because the normalized Laplacian $\mathbold{S}$ also changes when $\mathcal{G}$ grows.

Let $\lambda_1\leq \lambda_2\leq \cdots \leq \lambda_n$ be the eigenvalues of $\mathbold{S}$.  The spectrum of Laplacian matrices is well studied (see, e.g., \citealt{Chung1997}).
We have $0 \leq \lambda_i \leq 2$ for all $i$. We always have that $\lambda_1=0$, with  $\mathbold{\iota}_n$  as eigenvector. The number of zero eigenvalues $\lambda_i$ equals the number of connected components in $\mathcal{G}$.
 Hence, if $\mathcal{G}$ is connected, then $\lambda_2>0$ is the smallest non-zero eigenvalue of the normalized Laplacian. 
 
Our theory involves conditions on $\lambda_2$ and on the degree structure of the network through various harmonic means thereof.
$\lambda_2$ can be seen as a measure of global connectivity of $\mathcal{G}$. To see this we note that it can be linked to the Cheeger constant,
$$
C := \min_{U\in\left\{ U\subset V: \, 0<\sum_{i\in U} d_i \leq \sum_{i\notin U} d_i  \right\}}
\frac{\sum_{i\in U} \sum_{j\notin U } (\mathbold{A})_{ij}}{\sum_{i\in U} d_i} .
$$
The constant $C\in [0,1]$ measures how difficult it is to separate $\mathcal{G}$ into two 
disconnected components by removing edges from it.
The numerator in the definition of $C$ is the total weight of the removed edges, the denominator 
is the total degree in the smallest of the two components.
A larger value of $C$ implies a more strongly-connected graph,
and it is linked to $\lambda_2$ through the inequalities 
\begin{align} \label{CheegerInequalities}  
2 C \geq \lambda_2 \geq 1 - \sqrt{1-C^2} \geq \frac{1}{2}\, C^2 ,
\end{align}
which are due to \cite{FriedlandNabben2002}.
Thus, like the Cheeger constant, $\lambda_2$ is a measure of global connectivity of the graph $\mathcal{G}$. 
Our results below allow for $\lambda_2\rightarrow 0$ as $\mathcal{G}$ grows, and so cover situations where the graph becomes increasingly more sparse. We will give explicit rates on $\lambda_2$ for consistent estimation to be possible.

\begin{example}[Erd\H{o}s-R\'enyi graph]
\label{ex:ER}
Consider the \cite{ErdosRenyi1959} random-graph model, where edges between $n$ vertices are formed independently with probability $p_n$. The threshold on $p_n$ for $\mathcal{G}$ to be connected is $\ln (n) / n$ (\citealt{HoffmanKahlePaquette2013}). That is, if
$
p_n = c \, {\ln (n)} / {n} 
$
for a constant $c$, then, as $n \rightarrow \infty$, with probability approaching one, $\mathcal{G}$ is disconnected if $c<1$ and connected if $c>1$. In the former case, $\lambda_2\rightarrow 0$ while, in the latter case, $\lambda_2\rightarrow 1$, almost surely. 
\end{example}

\subsection{Two-way regression model on bipartite graph}
\label{sec:bipartite}

To relate our model to our main motivating examples consider the case of a bipartite graph $\mathcal{G}$, i.e., $V=V_1\cup V_2$ and $V_1\cap V_2 = \varnothing$, and edges are formed only between the subsets $V_1$ and $V_2$ but not within. So, for an edge $(i,j)$ we necessarily have that $i\in V_1$ and $j\in V_2$. A bipartite graph describes the interaction between two types of units, such as workers and firms or students and teachers. The outcome of interest here would typically be (log) wages or earnings and test scores, respectively. If we have panel data, so $\mathcal{G}$ is a multigraph, we may observe workers match with different firms over time and observe students in different classrooms or across multiple subjects. In fact, in these applications, such longitudinal data are necessary for $\mathcal{G}$ to be connected. A two-way regression model for such data takes the form
\begin{equation} \label{eq:twoway}
\mathbold{y} = \mathbold{B}_1\mathbold{\mu} + \mathbold{B}_2\mathbold{\eta} + \mathbold{X}\mathbold{\beta} + \mathbold{u},
\end{equation}
where $\mathbold{\mu}:=(\mu_1,\ldots,\mu_{n_1})^\prime$ and $\mathbold{\eta}:=(\eta_1,\ldots,\eta_{n_2})^\prime$ are the $n_1:=\lvert V_1 \rvert$ and $n_2:=\lvert V_2 \rvert$ parameter vectors for the two types of units,
and the $m\times n_1$ and $m\times n_2$ matrices $\mathbold{B}_1$ and $\mathbold{B}_2$ have entries
\begin{equation*}
\begin{split}	
(\mathbold{B}_1)_{\varepsilon_ei} 
& := 
\left\lbrace \begin{array}{rl}
1 & \text{if $e\in E_{(i,j)}$ for some $j\in V_2$}
\\
 0 & \text{otherwise,} 
\end{array}\right. 
\\
(\mathbold{B}_2)_{\varepsilon_ej} 
& := 
\left\lbrace \begin{array}{rl}
1 & \text{if $e\in E_{(i,j)}$ for some $i\in V_1$}
\\
 0 & \text{otherwise.} 
\end{array}\right. 
\end{split}
\end{equation*}
This is a workhorse specification to capture heterogeneity across units in linked data sets. It can be cast into \eqref{eq:regression1} by setting
$$
\alpha_i =
\left\lbrace
\begin{array}{rl} 
  \mu_i & \text{ if } i\in V_1 ,
\\
 -\eta_i & \text{ if } i\in V_2 ,
\end{array}
\right.
$$
sorting the units in $V$ by type so that we can write  $\mathbold{\alpha}=(\mathbold{\mu}^\prime,-\mathbold{\eta}^\prime)^\prime$, and constructing the $m\times n$ matrix $\mathbold{B}=(\mathbold{B}_1,-\mathbold{B}_2)$ by concatenation. 
Choosing the sign in front of $\eta_i$ is without loss of generality  because links are only formed between, but never within, the subsets $V_1$ and $V_2$. The need for a normalization built-in in our general specification arises here from the fact that \eqref{eq:twoway} is invariant to reparametrizations of the form $(\mu_i,\eta_j)\mapsto (\mu_i+c, \eta_j-c)$ for any $c$.

The two-way regression model provides an interesting example where a weighted graph arises naturally. In many applications the researcher is primarily interested in learning the parameters of one type, say those $i\in V_2$. This is so in teacher value-added models, for example. There, interest lies in estimating the $n_2$ teacher effects while controlling for unobserved student-specific heterogeneity through the inclusion of $n_1$ student effects (see, e.g., \citealt{JacksonRockoffStaiger2014}). 
Partialling-out the vector $\mathbold{\mu}$ from the two-way model in \eqref{eq:twoway} gives 
\begin{equation} \label{eq:profiled_eq}
\mathbold{M}_{\mathbold{B}_1} \mathbold{y}
=
(\mathbold{M}_{\mathbold{B}_1} \mathbold{B}_2)\, \mathbold{\eta} +
(\mathbold{M}_{\mathbold{B}_1}\mathbold{X})\mathbold{\beta}  +  \mathbold{M}_{\mathbold{B}_1}\mathbold{u},
\qquad\quad
\mathbold{M}_{\mathbold{B}_1}: = \mathbold{I}_m - \mathbold{B}_1(\mathbold{B}^\prime_1 \mathbold{B}_1)^\pinv \mathbold{B}^\prime_1
\, .
\end{equation}
From standard partitioned-regression theory, the least-squares estimator of $\mathbold{\eta}$ from this equation is numerically identical to the one obtained from joint estimation of $\mathbold{\mu}$ and $\mathbold{\eta}$ in \eqref{eq:twoway}. However, the formulation in \eqref{eq:profiled_eq} is helpful in understanding the behavior of the estimator of $\mathbold{\eta}$.
The properties of the matrix $\mathbold{B}_2^\prime\mathbold{M}_{\mathbold{B}_1}\mathbold{B}_2$ drive the sampling behavior of $\check{\mathbold{\eta}}$. This matrix is the Laplacian of a weighted one-mode projection \citep[p.~124]{Newman2010} of the bipartite graph $\mathcal{G}$ on the $n_2$ vertices in $V_2$. 

It is instructive to discuss this one-mode projection in more detail and to formalize how it fits the general setup in \eqref{eq:regression1}. 
Projecting the bipartite graph $\mathcal{G}=\mathcal{G}(V_1\cup V_2,E)$ on $V_2$ is done by suppressing the vertices in $V_1$. This gives a new (unipartite) graph, say $\mathcal{G}^\prime=\mathcal{G}(V_2,E^\prime)$. Each edge pair $(e_1,e_2)$ with $e_1\in E_{(i,j)}$ and $e_2\in E_{(i,j^\prime)}$ in $\mathcal{G}$ for some $i\in V_1$ and $j, j^\prime\in V_2$ gives rise to a single edge $e=(e_1,e_2)\in E^\prime_{(j,j^\prime)}$ in $\mathcal{G}^\prime$. In the student-teacher example, two teachers $j$ and $j^\prime$ are connected by an edge in $\mathcal{G}^\prime$ if and only if there exists at least one student $i$ that they have both taught. 
Alternatively, the edge $e=(e_1,e_2)\in E^\prime$ exists because $e_1\in E$ and $e_2\in E$ both connect to the same vertex $i\in V_1$. Given the edges $e_1,e_2$ this connecting vertex $i$ is unique; for later use we denote it by $c((e_1,e_2))$.
In $\mathcal{G}^\prime$ we have
$$
m^\prime := \lvert E^\prime \rvert = \sum_{i\in V_1} \frac{d_i(d_i-1)}{2}
$$
edges; $m^\prime$ need not equal $m$ and, indeed, may be much larger. 
We again label $e^\prime\in E^\prime$ by natural numbers $\varepsilon^\prime_{e^\prime}$.
The process of concatenating edges in $E$ to form the new edge set $E^\prime$ can be described by the $m^\prime\times m$ matrix $\mathbold{Q}$ with entries
$$
(\mathbold{Q})_{\varepsilon^\prime_{e^\prime} \, \varepsilon_{e_1}}
:=
\left\lbrace
\begin{array}{rl}
  1 & \text{ if } e^\prime=(e_1,e_2) \text{ for some } e_2\in E,  \\
 -1 & \text{ if } e^\prime=(e_2,e_1) \text{ for some } e_2\in E,  \\
  0 & \text{otherwise.}
\end{array}\right.	
$$
Choosing the orientation of the rows of $\mathbold{Q}$ is without loss of generality. The matrix $\mathbold{Q}$ has a first-differencing interpretation. Indeed, when applied to the two-way regression model \eqref{eq:twoway} we get
$
\mathbold{Q}\mathbold{y} = \mathbold{Q}\mathbold{B}_2\mathbold{\eta} + \mathbold{Q}\mathbold{X}\mathbold{\beta} + \mathbold{Q}\mathbold{u} ,
$
because $\mathbold{Q}\mathbold{B}_1 = 0$.
Thus, $\mathbold{Q}$ sweeps out the $n_1$ nuisance parameters $\mathbold{\mu}$ and transforms original outcomes into first differences. The matrix $\mathbold{Q}\mathbold{B}_2$ is the (oriented) incidence matrix of an unweighted graph, and this first-differenced regression equation fits \eqref{eq:regression1}.\footnote{%
$\mathbold{Q}\mathbold{B}_2$ will contain rows with only zero entries if there are differenced outcomes that do not depend on $\mathbold{\eta}$. This is at odds with the definition of an incidence matrix. Dropping these differences from $\mathbold{Q}\mathbold{y}$, however, restores the incidence matrix interpretation of $\mathbold{Q}\mathbold{B}_2$. This operation does not affect estimation of $\mathbold{\eta}$ and so is irrelevant for our purposes. However, the differenced outcomes may provide information on $\mathbold{\beta}$, which is why we prefer to work with $\mathbold{Q}\mathbold{B}_2$ as defined here.
}
Applying least-squares directly to the first differences is inefficient and is not equivalent to estimation of the two-way regression model.  
Ordinary least-squares estimation of \eqref{eq:profiled_eq} is numerically equivalent to weighted least-squares estimation of the first-differenced equation. The relevant $m^\prime\times m^\prime$ diagonal weight matrix $\mathbold{W}$ has entries
$
(\mathbold{W})_{\varepsilon^\prime_e\, \varepsilon^\prime_e} := {1}/\sqrt{d_{c(e)}}.
$
Ordinary least-squares applied to \eqref{eq:profiled_eq} and to 
\begin{equation} \label{eq:profiled_eq2}
{\mathbold{W}}\mathbold{Q}\mathbold{y}
=
{\mathbold{W}}\mathbold{Q}\mathbold{B}_2\mathbold{\eta}
+
{\mathbold{W}}\mathbold{Q} \mathbold{X}\mathbold{\beta}
+
{\mathbold{W}}\mathbold{Q} \mathbold{u}
\end{equation}
yields the same result.
Here, ${\mathbold{W}}\mathbold{Q}\mathbold{B}_2$ is the incidence matrix of a weighted one-mode projection of $\mathcal{G}$. This $\mathcal{G}^\prime$ determines the properties of the least-squares estimator. Its Laplacian is 
$$
\mathbold{L}^\prime
:= 
\mathbold{B}_2^\prime 
(\mathbold{Q}^\prime \mathbold{W}^2 \mathbold{Q})
\mathbold{B}_2
=
\mathbold{B}_2^\prime\mathbold{M}_{\mathbold{B}_1}\mathbold{B}_2,
$$ 
where we use the fact that $
\mathbold{Q}^\prime \mathbold{W}^2 \mathbold{Q}=\mathbold{M}_{\mathbold{B}_1}$.

The adjacency matrix of $\mathcal{G}^\prime$ is the $n_2\times n_2$ matrix ${\mathbold{A}}^\prime$ with entries
\begin{equation*} 
({\mathbold{A}}^\prime)_{j j^\prime} :=
\left\lbrace
\begin{array}{cl}
\displaystyle
\sum_{i \in [j] \cap [j^\prime]}  \frac{\lvert E_{(i,j)} \rvert  \, \lvert E_{(i,j^\prime)} \rvert } {\sum_{k\in V_2} \lvert E_{(i,k)} \rvert}  & \text{ for $j \neq j^\prime$,}	
\\[5pt]
0 & \text{  for $j=j^\prime$.}
\end{array}	
\right. 
\end{equation*}
Here, $[j] \cap [j^\prime]$ is the set of all vertices in $V_1$ that are connected to both $j\in V_2$ and $j^\prime \in V_2$ in the original bipartite graph $\mathcal{G}$. In the student-teacher example two teachers are connected by an edge if there is at least one student who was taught by both teachers. The weight $({\mathbold{A}}^\prime)_{jk}$ of the edge is larger the more students there are connecting teachers $j$ and $k$, and the more courses they have taken from these teachers. $\mathcal{G}^\prime$ determines the accuracy with which teacher value-added can be estimated.

\begin{figure}
\centering
\begin{subfigure}{.5\textwidth}
  \centering
  \includegraphics[width=.6\textwidth]{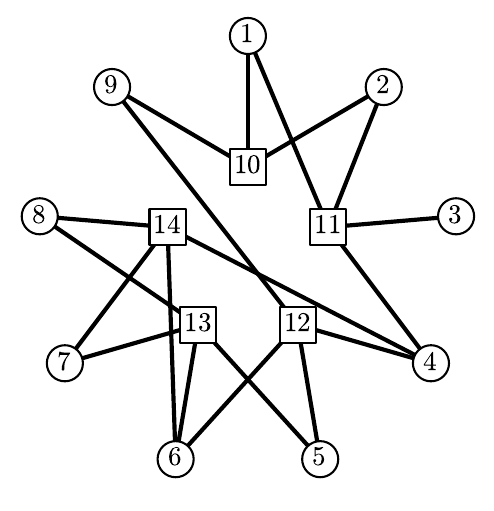}
\end{subfigure}%
\begin{subfigure}{.5\textwidth}
  \centering
  \includegraphics[width=.6\textwidth]{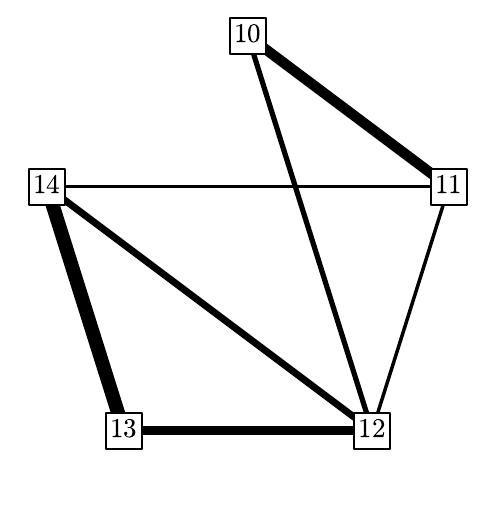}
\end{subfigure}
\caption{ \footnotesize
A simple unweighted bipartite graph (left) with links between $V_1$ (circular vertices) and $V_2$ (square vertices),
 and the induced weighted graph (right) on $V_2$ alone resulting from profiling out the parameters associated with $V_1$. }
\label{fig:weighted}
\end{figure}

The matrix $\mathbold{A}^\prime$ is also the adjacency matrix of the simple graph obtained from $\mathcal{G}^\prime$ by replacing all edges $e\in E_{(j,j^\prime)}^\prime$ by one weighted edge, with weight $(\mathbold{A}^\prime)_{jj^\prime}$.
Figure \ref{fig:weighted} provides an illustration of a simple bipartite graph for students (circular vertices) and teachers (square vertices), given in the left plot, and its induced weighted graph featuring only teachers, given in the right plot. The thickness of the edge between $(j,j^\prime)$ in the latter plot reflects the magnitude of the weight $({\mathbold{A}}^\prime)_{jj^\prime}$. 

The device of a one-mode projection highlights the importance of having movers in panel data. 
In matched worker-firm data sets workers do not frequently switch employer over the course of the sampling period. This lack of mobility is one cause of the substantial bias that is observed in the correlation coefficient between (estimated) worker and firm effects (\citealt{AbowdKramarzLengermannPerezDuarte2004}, \citealt{AndrewsGillSchankUpward2008,AndrewsGillSchankUpward2012}). While this is now well recognized, limited mobility has consequences more broadly. Indeed, it implies that few workers connect firms in the one-mode firm projection. Therefore, the induced graph may be only weakly connected (and $\lambda_2$ will be close to zero) and the variance of the estimator of the firm effects may be large. This is not only detrimental for identifying sorting between workers and firms but, indeed, complicates estimation and inference of the firm effects as well as all their moments, such as their variance. Restricting attention to large firms need not resolve this problem. An analogous argument holds for teacher effects and their estimated variance, and so for our ability to infer the contribution of teacher value-added to observed variation in test scores. We illustrate this in our data below.

\section{Variance bound and asymptotic analysis}

\subsection{Finite-sample bound}

To work towards general distribution theory it is instructive to start with a finite-sample bound on the variance of the fixed-effect estimator when the errors are homoskedastic and uncorrelated.
Let
\begin{align}
h_i := \left( \frac{1}{d_i} \sum_{j \in [i]} \frac{(\mathbold{A})_{ij}^2}{d_j} \right)^{-1}.
   \label{DefHarmonicMean}
\end{align}
This is a (weighted) harmonic mean of the (weighted) degrees $d_j / (\mathbold{A})_{ij}$ of all $j\in [i]$. Note that, for a given vertex $i$, $h_i$ is increasing in the degree of its direct neighbors.

\begin{theorem}[Variance bound] \label{thm:firstorderbound}
Let $\mathcal{G}$ be connected. Suppose that $\mathbold{u}\sim(\boldsymbol{0},\sigma^2 \mathbold{I}_m)$. Then
\begin{align*}
 \frac{\sigma^2} {d_i}  - \frac{2\, \sigma^2} {m}  
 \,   \leq    \,  {\rm var}( \hat{\alpha}_i  )  \,     \leq   \,
   \frac{\sigma^2} {d_i}   \left( 1  +  \frac{1}  {\lambda_2 h_i} \right)  - \frac{2\, \sigma^2} {m}   .
\end{align*}
\end{theorem}

\noindent
Theorem \ref{thm:firstorderbound} states that, for a given degree $d_i$ and global connectivity measure $\lambda_2$, the upper bound on the variance of $\hat{\alpha}_i$ is smaller if the direct neighbors of vertex $i$ are themselves more strongly connected to other vertices in the network.
The theorem provides insight into how the local connectivity structure of the network, around vertex $i$, affects statistical precision.

\begin{ContExample}{\ref{ex:ER}}%
Consider the \cite{ErdosRenyi1959}  random-graph model with $p_n = c \, \ln (n) / n$ for $c>1$. Let $i$ be a randomly chosen vertex. Then, as $n \rightarrow \infty$, we have, almost surely,
$$
\lambda_2\rightarrow 1, \qquad
\frac{d_i}{\ln n} \rightarrow c, \qquad \frac{h_i}{\ln n} \rightarrow c .
$$
Consequently, 
$$
{\rm var}( \hat{\alpha}_i  ) = \frac{\sigma^2}{d_i} + O(d^{-2}_i)
$$
follows from Theorem \ref{thm:firstorderbound}.
\end{ContExample}

\noindent
Additional calculations for analytically-tractable cases where $\lambda_2 \rightarrow 0$ as the network grows are provided in the supplementary material.

Theorem \ref{thm:firstorderbound} highlights the importance of $\lambda_2 h_i \rightarrow \infty$ as a sufficient condition
for the parametric rate ${d_i}^{-1/2}$ to be attainable for estimation of $ \hat{\alpha}_i$. This result carries over to the model with covariates. Let
$$
\rho : = 
\left\lVert (\mathbold{X}^\prime \mathbold{X})^{-1/2}  
(\mathbold{X}^\prime \mathbold{M}_{\mathbold{B}} \mathbold{X} )
 (\mathbold{X}^\prime \mathbold{X})^{-1/2} \right\rVert_2, 
$$
where $\lVert \cdot \rVert_2$ denotes the spectral norm, and
$\mathbold{M}_{\mathbold{B}}: = \mathbold{I}_m - \mathbold{B}(\mathbold{B}^\prime \mathbold{B})^\pinv \mathbold{B}^\prime$. 
 Note that  $\rho \in [0,1]$ is a measure of non-collinearity between the columns of $\mathbold{X}$ and $\mathbold{B}$, with $\rho$ close to zero indicating near-collinearity. Indeed, while $\mathbold{X}^\prime  \mathbold{X}$ measures the total variation in $\mathbold{X}$, 
$\mathbold{X}^\prime \mathbold{M}_{\mathbold{B}} \mathbold{X}$ captures the residual variation in $\mathbold{X}$ after its linear dependence on $\mathbold{B}$ has been partialled out.
For $i \in V$ let $\mathbold{b}_i$ be $i^{\text{th}}$ column of $\mathbold{B}$, and let
$
\overline{\mathbold{x}}_i : =  {\mathbold X}' \mathbold{b}_i / d_i
$
and
$\mathbold{\Omega}:= \mathbold{X}^\prime \mathbold{X}/m$ in the following theorem.

\begin{theorem}[Variance bound (cont'd.)] \label{thm:regressors}
Let $\mathcal{G}$ be connected. Suppose that $\mathbold{u}\sim(\boldsymbol{0},\sigma^2 \mathbold{I}_m)$, $\mathrm{rank}(\mathbold{X})=p$, and  $\mathrm{rank}((\mathbold{X},\mathbold{B}))=p+n-1$. Then
\begin{align*}
\left|
{\rm var}\left(\check \alpha_i \right) - {\rm var}\left(\hat \alpha_i \right)
\right|
\; \leq  \;      
\frac{2 \, \sigma^2} \rho
\left(     
\frac {1-\rho} {d_i \, (\lambda_2\, h_i)}   
+
\frac { \overline {\mathbold x}'_i   \, {\mathbold \Omega}^{-1}   \overline {\mathbold x}_i } m
\right)  ,
\end{align*}
for all $i\in V$.
\end{theorem}

\noindent
This result shows that, if $\rho$ is bounded away from zero, introducing covariates only has a higher-order effect on the statistical precision of the fixed-effect estimator. In particular we have 
\begin{align}
    {\rm var}( \check{\alpha}_i  ) = \frac{\sigma^2}{d_i} + o(d^{-1}_i),
    \label{FirstOrderRate}
\end{align}
provided that $\lambda_2 h_i \rightarrow \infty$ as $\mathcal{G}$ grows. Furthermore, the parametric rate is achievable even if $\lambda_2$ is not treated as fixed, and $\mathcal{G}$ becomes less dense as more vertices are added to the network.

\subsection{Large-sample analysis}
\label{sec:LargeSample}

We now discuss asymptotic results under more general conditions on the regression errors. 
The following theorem provides a first-order representation of $ \check{\alpha}_i$.
Let $\mathbold{\Sigma}:=\mathbbmsl{E}(\mathbold{u}\mathbold{u}^\prime)$.

\begin{theorem}[First-order representation] \label{thm:firstorderGeneralized}
Let $\mathcal{G}$ be connected.
Assume that $\mathrm{rank}(\mathbold{X})=p$ and $\mathrm{rank}((\mathbold{X},\mathbold{B}))=p+n-1$.
Suppose that $\mathbbmsl{E}(\mathbold{u})=\boldsymbol{0}$ and that $\lVert \mathbold{\Sigma}\rVert_2 \leq \overline{\sigma}^2 = O(1)$. 
Then
$$
 \check{\alpha}_i-\alpha_i = \frac{\mathbold{b}^\prime_i \mathbold{u}}{d_i}  + \epsilon_i + \tilde \epsilon_i  \, ,
$$
where $\epsilon_i$ and $\tilde\epsilon_i $ are zero-mean random variables that satisfy
$\mathbbmsl{E}(\epsilon_i^2)\leq \overline{\sigma}^2 (1+\rho) /(\rho \, d_i \, \lambda_2 \, h_i)$,
and $\mathbbmsl{E}(\tilde \epsilon_i^{\,2})\leq 
     \overline{\sigma}^2  \,   \overline{\mathbold{x}}_i'  \, \mathbold{\Omega}^{-1} \, \overline{\mathbold{x}}_i  / (\rho \, m) $.  
\end{theorem}

\noindent
From the definition of the incidence matrix in \eqref{eq:incidence}, the $m$-vector $\mathbold{b}_i$ has as many non-zero entries as there are edges involving vertex $i$. Further, $\mathbold{b}_i^\prime \mathbold{b}_i = d_i$. Hence, the term ${\mathbold{b}^\prime_i \mathbold{u}} /{d_i}$ is a (weighted) sample mean of the regression errors associated with the edges that involve vertex $i$.

We next consider sequences of growing networks such that
\begin{align} \label{AsymptoticConditionsBeta} 
\rho^{-1} = O(1), &&
d_i / m \rightarrow 0,  &&
\overline{\mathbold{x}}_i'  \, \mathbold{\Omega}^{-1} \, \overline{\mathbold{x}}_i = O(1).
\end{align}
These are relatively weak conditions that ensure that the fact that $\mathbold{\beta}$ is estimated can be ignored in large samples. Moreover, they imply that
$$
\epsilon_i = O_p\left(\frac{1}{\sqrt{d_i (\lambda_2\, h_i)}}\right)
$$
and that $\tilde \epsilon_i  = O_p(1/\sqrt{m}) = o_p(1/\sqrt{d_i})$.
The main implication of the theorem is that, then, under the now familiar condition $\lambda_2 h_i\rightarrow\infty$,
 as $d_i\rightarrow\infty$,
\begin{align*}
(\check{\alpha}_i-\alpha_i )
\; \overset{p}{\rightarrow} \;
\frac{ \mathbold{b}^\prime_i \mathbold{u}} {d_i} .
\end{align*}
This result allows the errors to be heteroskedastic and correlated.

With Theorem \ref{thm:firstorderGeneralized} in hand the limit distribution of $\check{\alpha}_i$ can be deduced under conventional conditions. As an example we do so next  for independent but heterogeneously distributed (i.n.i.d.) regression errors. 

\begin{theorem}[Limit distribution for i.n.i.d.~errors] \label{thm:inid}
Let the assumptions of Theorem \ref{thm:firstorderGeneralized} 
and the conditions in \eqref{AsymptoticConditionsBeta} hold.
Suppose that the regression errors are independent,
 have bounded fourth-order moments,
and variances bounded away from zero,
and that the edge weights are bounded away from zero and from infinity. Then
$$ 
\frac{\check{\alpha}_i - \alpha_i}
{\sqrt{\mathbold{b}_i^\prime\mathbold{\Sigma}\mathbold{b}_i}/d_i}
\overset{d}{\rightarrow} N(0,1),
$$
as $d_i\rightarrow\infty$, provided that $\lambda_2 h_i\rightarrow\infty$.
\end{theorem}

\noindent
When the errors $\mathbold{u}$ are independent and homoskedastic we have  $\mathbold{b}_i^\prime{\mathbold{\Sigma}}\mathbold{b}_i = \sigma^2\, \mathbold{b}_i^\prime\mathbold{b}_i = \sigma^2  d_i$ and the variance in the theorem reduces to $\sigma^2/d_i$, which agrees with \eqref{FirstOrderRate}.

A plug-in estimator of $\mathbold{b}_i^\prime{\mathbold{\Sigma}}\mathbold{b}_i$ is
$\mathbold{b}_i^\prime\check{\mathbold{\Sigma}}\mathbold{b}_i$, where
$\check{\mathbold{\Sigma}}:= \mathrm{diag}(\check{\mathbold{u}}\check{\mathbold{u}}^\prime)/{m}$ and $\check{\mathbold{u}}$ are the residuals from the least-squares regression. This involves estimation of $\alpha_j$ for all $j\in[i]$. We have that
$$
(\mathbold{b}_i^\prime\check{\mathbold{\Sigma}}\mathbold{b}_i
-
\mathbold{b}_i^\prime{\mathbold{\Sigma}}\mathbold{b}_i)/d_i
\overset{p}{\rightarrow} 0 
$$
as $d_i\rightarrow \infty$, provided that, in addition to the conditions of Theorem \ref{thm:inid} holding, we have that $\lambda_2 H_i$ is bounded away from zero, where
$$
H_i : = \left( \sum_{j\in[i]} \frac{(h_i/d_i)/d_j}{h_j} \right)^{-1}
$$
is a weighted harmonic mean. At the heart of this result lies (a local version of) a global convergence rate on ${\lVert \check{\mathbold{\alpha}}-\mathbold{\alpha} \rVert}$, which is interesting in its own right. More precisely, letting
$$
h:= \left(\frac{1}{n} \sum_{i=1}^n \frac{1}{d_i} \right)^{-1}
\qquad \text{ and } \qquad
H : = \left( \sum_{i=1}^n \frac{(h/n)/d_i}{h_i} \right)^{-1},
$$
it is easy to see that 
$$
{\lVert \check{\mathbold{\alpha}}-\mathbold{\alpha} \rVert} = 
O_p({\textstyle{\sqrt{n/h}}}),
$$
provided that $\lambda_2 H$ is bounded away from zero.

\subsection{Estimation of moments}

Suppose that the $\alpha_i$ are sampled from some distribution. One might be interested to learn the variance of this distribution---as in, say, \cite{Rockoff2004} or \cite{CardHeiningKline2013}---or some other moment. The typical estimator  is the corresponding sample moment of the estimated effects. Sampling noise in the estimated individual effects will introduce bias in the moment estimator, however. To see this, consider estimation of the variance in a simple model without regressors. The sample variance of the estimated effects in this case is
$$
{\hat{\mathbold{\alpha}}^\prime \mathbold{M}_{\mathbold{\iota}_n} \hat{\mathbold{\alpha}}}/{(n-1)},
$$
where 
$
\mathbold{M}_{\mathbold{\iota}_n}
:= 
\mathbold{I}_n -{\mathbold{\iota}_n\mathbold{\iota}_n^\prime}/{n}$ is the usual demeaning matrix. When $\mathbold{u}\sim(\boldsymbol{0},\sigma^2 \mathbold{I}_m)$ its bias is
$$
\sigma^2 \, \frac{\mathrm{tr}(\mathbold{L}^\pinv)}{(n-1)} = (n-1)^{-1} \sum_{i=1}^n \mathrm{var}(\hat{\alpha}_i),
$$
which clearly shows how imprecise estimation of $\alpha_i$ contributes to the bias in the variance estimator.

It is difficult to derive an exact expression for the bias for more general functionals. Theorem~\ref{thm:inid} is instrumental here. Suppose that  $\tau:=\mathbbmsl{E}(\varphi(\alpha_i))$ is of interest.  Its plug-in estimator is
$$
\check{\tau}:= n^{-1}\sum_{i=1}^n \varphi(\check{\alpha}_i).
$$
Under the conditions of the theorem we can calculate the leading bias in this estimator as
$$
b: = n^{-1} \sum_{i=1}^n {\mathbbmsl{E}\left(\frac{\varphi^{\prime\prime}(\alpha_i)}{2} \, \frac{\mathbold{b}_i^\prime\mathbold{\Sigma}\mathbold{b}_i/d_i}{d_i}\right)}
= O(h^{-1}),
$$
where $\varphi^{\prime\prime}$ denotes the second derivative, provided $\lambda_2 \, H \rightarrow  \infty$. Simple regularity conditions on $\varphi$ for this bias result to hold are that it is differentiable with $\mathbbmsl{E}(\varphi^{\prime\prime}(\alpha_i)^2) < \infty$ and bounded third derivative. So, quite generally, the bias will shrink like $h^{-1}$. Therefore, for the bias to vanish and $\check{\tau}$ to be consistent, we need that the degrees of the individual vertices grow with $n$ for an increasing fraction of the vertices.

If the functional of interest is the variance, an exact bias correction can be performed (see  \citealt{AndrewsGillSchankUpward2008} and \citealt{KlineSaggioSolvsten2018}). For  functionals like $\tau$, a plug-in estimator of the leading-order bias $b$ is easily formed and so an adjusted estimator is readily constructed. Its effectiveness as a bias-correction device will again depend on the connectivity structure of the graph.
We postpone a detailed analysis to future work. In a recent contribution \cite{KlineSaggioSolvsten2018} present limit theory for quadratic forms in $\mathbold{\alpha}$.

\section{Empirical illustrations}

\subsection{Teacher value-added}
We construct a graph connecting teachers as the (weighted) one-mode projection from matched student-teacher data from the North Carolina Education Research Center. The projection of interest is the one discussed in Section 2.3. The full data set includes scores for a standardized test in reading in elementary schools in North Carolina and was used by \cite{Verdier2016} to estimate the effect of class-size reduction on student performance. The analysis conducted here is useful to assess the precision with which teacher value-added can be estimated. The data concern pupils in Grades 4 and 5 of elementary school over the period 2008--2012. The full teacher graph (with a single weighted edge between neighboring teachers, as in Figure \ref{fig:weighted}) has 12,057 vertices and 53,741 edges and is disconnected. The largest connected component involves 41,612 edges between 11,945 teachers and we work with this subgraph.
With $\lambda_2=.0039$ the projected teacher graph is weakly connected. 
Its local connectivity is summarized in Table \ref{tab:teacherdistributions}. The table contains the mean, standard deviation, and deciles of the relevant degree distributions.
Inspection reveals that the degrees are small for all teachers.

\begin{table}[h!] \footnotesize
  \centering
  \caption{Summary statistics for the teacher graph}
 \resizebox{\columnwidth}{!}{ 
    \begin{tabular}{crrrrrrrrrrr}
		\hline\hline
          & \multicolumn{1}{l}{mean} & \multicolumn{1}{l}{stdev}  & \multicolumn{1}{c}{$10^{\mathrm{th}}$\%}     & \multicolumn{1}{c}{$20^{\mathrm{th}}$\%}     & \multicolumn{1}{c}{$30^{\mathrm{th}}$\%}     & \multicolumn{1}{c}{$40^{\mathrm{th}}$\%}     & \multicolumn{1}{c}{$50^{\mathrm{th}}$\%}     & \multicolumn{1}{c}{$60^{\mathrm{th}}$\%}       & \multicolumn{1}{c}{$70^{\mathrm{th}}$\%}       & \multicolumn{1}{c}{$80^{\mathrm{th}}$\%}       & \multicolumn{1}{c}{$90^{\mathrm{th}}$\%}        \\
		  \hline
    $d_i$    & 13.87 & 10.76   & 3.00  & 5.50  & 7.50 & 9.00  & 11.00 & 14.00 & 17.50 & 21.50 & 27.50  \\
    $h_i$    & 7.15 & 7.13   & 2.43  & 3.30  & 4.01  & 4.72 & 5.48 & 6.36 & 7.44 & 9.12 & 12.56  \\
    $H_i$    & 36.48 & 58.59   & 3.03  & 5.72  & 10.48 & 14.76 & 19.81 & 26.20 & 35.67 & 50.65 & 83.48  \\
	\hline\hline
    \end{tabular}
	}
  \label{tab:teacherdistributions}
\end{table}

\begin{figure}[h!]
\centering
  \centering
  \includegraphics[width=1\textwidth]{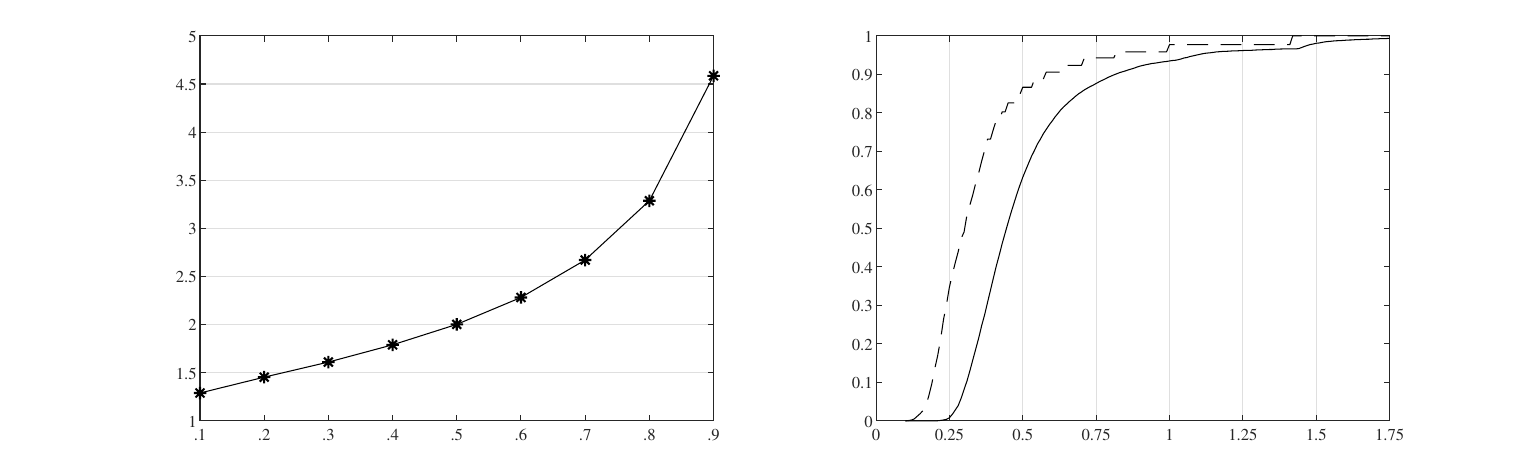}
\caption{ \footnotesize
Deciles of the distribution of $(\mathbold{S}^\dagger)_{ii}$ (left plot) and empirical distributions of the width of $95\%$ confidence bands (right plot). The width is calculated as $2\times(1.96 \,(\mathbold{L}_{ii}))$ (solid curve) and $2\times(1.96 \, d_{i}^{-1})$ (dashed curve).
}
\label{fig:approximations}
\end{figure}

The weak connectivity suggests that inference on teacher value-added will be difficult. To get a sense of the precision of a first-order asymptotic approach we can look at the ratio
$$
\frac{(\mathbold{L}^\pinv)_{ii}}{1/d_i} = (\mathbold{S}^\dagger)_{ii}.
$$
This is the exact variance of $\hat{\alpha}_i$ to its large-sample approximation in a regression model with homoskedastic and uncorrelated errors. This ratio is free of $\sigma^2$ and can be computed directly from the graph. The left plot of Figure \ref{fig:approximations} shows the deciles of the distribution of $(\mathbold{S}^\dagger)_{ii}$. The asymptotic approximation is revealed to be widely inaccurate. On average, the actual variance is about 2.5 times larger than its approximation. Even the first decile equals $1.29$. This implies that confidence intervals based on the large-sample arguments in Theorem~\ref{thm:firstorderGeneralized} are overoptimistic. 
To illustrate this the right plot in Figure \ref{fig:approximations} gives the distribution of the width of 95\% confidence intervals for the $\alpha_i$ using both the exact variance (solid line) and its large-sample approximation (dashed line) for the case $\sigma^2=1$. The former stochastically dominates the latter.

The large variability in the estimators of teacher value-added implies a large bias in their estimated variance. We calculate
$$
\mathrm{tr}(\mathbold{L}^\pinv)/(n-1) = .3545,
$$
so the bias in the plug-in estimator of the variance is about one-third of the error variance when $\mathbold{u}\sim(\boldsymbol{0},\sigma^2\mathbold{I}_m)$. The large-sample approximation to the bias here is proportional to $h^{-1}$. With $h=5.4554$ this yields a bias of about 18\%, roughly half the size of the exact bias.

One reason for the global connectivity of the teacher graph to be low is limited mobility of teachers between schools. If interest lies in comparing teacher effectiveness within a given school it suffices to restrict attention to that subgraph. Of course, the effective sample size from which teacher value-added is estimated will remain small unless additional years of data are collected. Making accurate comparisons between schools is more complicated as it requires many teachers teachers to switch schools during the sampling period. Collecting additional years of data will not automatically lead to more precise estimates. \cite{Mansfield2015} discusses the feasibility of ranking teachers within and between schools (see also \citealt{MihalyMcCaffreySassLockwood2013} for related discussions). A previous draft of this paper contains versions of our main theoretical results specialized to within and between decompositions of graphs.

\subsection{Occupational network}
Wage regressions on worker and occupational dummies (as in \citealt{KambourovManovskii2009}, for example) provide an interesting example of a situation where more accurate results can be obtained. We use all 18 available waves from the BHPS (for a total of 132,097 observations) to construct the induced (weighted) occupational network. The Standard Occupational Classification (SOC90) in the BHPS distinguishes (at the three-digit level) between 374 occupations. 
We again focus on the largest connected component, which contains 365 occupations with 14,825 weighted edges between them.
As a measure of global connectivity here we find $\lambda_2=.3289$. Compared to traditional matched employer-employee data our occupational network does not suffer as much from limited mobility. One reason is that the number of occupations is relatively small compared to the number of workers. Another is that workers may switch occupation also if they remain employed by the same firm, for example due to internal promotions. Finally, as we are dealing with self-reported occupations there is also the possibility of spurious mobility due to misreporting. A look at the distributions summarized in Table \ref{tab:occdistributions} reveals that the degrees and harmonic means tend to be larger here than in the teacher graph.

\begin{table}[htbp] \footnotesize
  \centering
  \caption{Summary statistics for the occupation graph}
 \resizebox{\columnwidth}{!}{ 
    \begin{tabular}{crrrrrrrrrrr}
		\hline\hline
          & \multicolumn{1}{l}{mean} & \multicolumn{1}{l}{stdev}  & \multicolumn{1}{c}{$10^{\mathrm{th}}$\%}     & \multicolumn{1}{c}{$20^{\mathrm{th}}$\%}     & \multicolumn{1}{c}{$30^{\mathrm{th}}$\%}     & \multicolumn{1}{c}{$40^{\mathrm{th}}$\%}     & \multicolumn{1}{c}{$50^{\mathrm{th}}$\%}     & \multicolumn{1}{c}{$60^{\mathrm{th}}$\%}       & \multicolumn{1}{c}{$70^{\mathrm{th}}$\%}       & \multicolumn{1}{c}{$80^{\mathrm{th}}$\%}       & \multicolumn{1}{c}{$90^{\mathrm{th}}$\%}        \\
		  \hline
    $d_i$    & 155.06 & 268.82 & 7.35 & 16.08 & 27.68 & 45.68 & 67.10 & 92.66 & 143.99 & 212.09 & 402.01 \\
	 $h_i$    & 81.50 & 125.20 & 14.52 & 20.88 & 27.25 & 35.16 & 46.23 & 60.61 & 80.72 & 113.67 & 163.74  \\
    $H_i$    & 213.77 & 455.93 & 11.15 & 22.90 & 35.29 & 48.59 & 66.34 & 102.82 & 172.75 & 296.80 & 539.07 \\
	\hline\hline
    \end{tabular}
	}
  \label{tab:occdistributions}
\end{table}

The distribution of $(\mathbold{S}^\dagger)_{ii}$ now places most of its mass in close vicinity of unity. Its mean and standard deviation are $1.034$ and $.0521$. The median is $1.0202$ while the first and ninth decile are $1.0046$ and $1.0759$, respectively. This suggests that, here, the large-sample approximation to the variance is a much more accurate reflection of actual estimation uncertainty. Similarly, we may again calculate
$\mathrm{tr}(\mathbold{L}^\pinv)/(n-1) = .0577$, which is about 7 times smaller than in the previous example. Further, as $h^{-1} = .0566$, here, the bias approximation is quite accurate.

\section{Conclusion}

We have presented inference results on individual effects in a linear fixed-effect regression model when the underlying 
data structure constitutes a (weighted) graph. An important example is a two-way regression model on a bipartite graph. The main contribution of this paper is to quantify the dependence of statistical precision of the estimator on
the connectivity structure of the graph. A key measure of global connectivity is the smallest non-zero eigenvalue of the (normalized) Laplacian matrix of the graph. It reflects the intuitive notion of mobility in the network. A small eigenvalue captures the presence of bottlenecks, which is detrimental to statistical precision. Several measures of local connectivity, such as the degree structure and various harmonic means thereof, also arise naturally in our analysis. 

Our theoretical work highlights the importance of and the interplay between global and local measures of network connectivity for conventional inferential approaches to be reliable. The analysis points to a set of simple statistics that can be inspected to evaluate whether the network is sufficiently well connected in a given application. In an application to teacher value-added we find that this is not the case. We further find that conventional standard errors on teacher value-added estimates are much too small, 
resulting in a false sense of (statistical) precision on these parameter estimates. In an occupational network, on the other hand, we find much higher measures of connectivity and support for our large-sample approximations.

\small

\newtheorem{theoremS}{Theorem}
\newtheorem{lemmaS}[theoremS]{Lemma}
\newtheorem{corollaryS}[theoremS]{Corollary}
\newtheorem{exampleS}{Example}
 
 \renewcommand{\thesection}{S.\arabic{section}}  \setcounter{section}{0}
\renewcommand{\theequation}{S.\arabic{equation}}  \setcounter{equation}{0}
\renewcommand{\thetheoremS}{S.\arabic{theoremS}}  \setcounter{theoremS}{0}
\renewcommand{\theexampleS}{S.\arabic{exampleS}}  \setcounter{exampleS}{0}
\renewcommand{\thefigure}{S.\arabic{figure}}  \setcounter{figure}{0}

\renewcommand{\qedsymbol}{$\blacksquare$}

\clearpage

\setcounter{section}{0}
\setcounter{footnote}{0}
\setcounter{page}{1}
\pagenumbering{roman}

\begin{center}
{\bf \large 
SUPPLEMENTARY MATERIAL FOR \\ `FIXED-EFFECT REGRESSIONS ON NETWORK DATA'} 
\end{center}

\section{Additional illustrations}

Recall that our measure of global connectivity of the graph $\mathcal{G}$ is $\lambda_2$, the second smallest eigenvalue of the normalized Laplacian matrix. In the following we provide some concrete examples of graphs for which $\lambda_2$ can be explicitly calculated, and we discuss the implications of our variance bound in Theorem \ref{thm:firstorderbound}

Our first example illustrates that, even if $\lambda_2\rightarrow 0$ with the sample size, we may still have that $\mathrm{var}(\hat{\alpha}_i) \asymp d_i^{-1}$.

\begin{figure}
\centering
\begin{subfigure}{.5\textwidth}
  \centering
  \includegraphics[width=.7\textwidth]{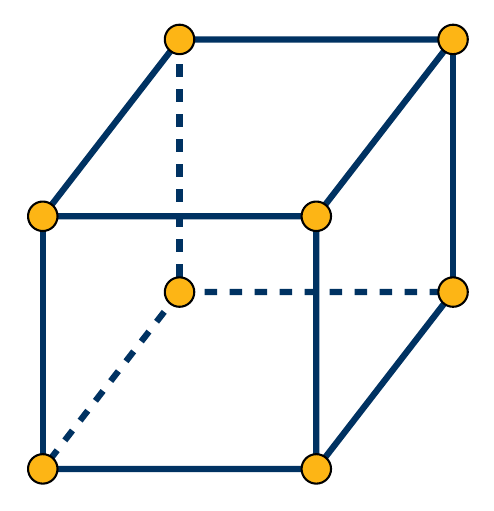}
\end{subfigure}%
\begin{subfigure}{.5\textwidth}
  \centering
  \includegraphics[width=.7\textwidth]{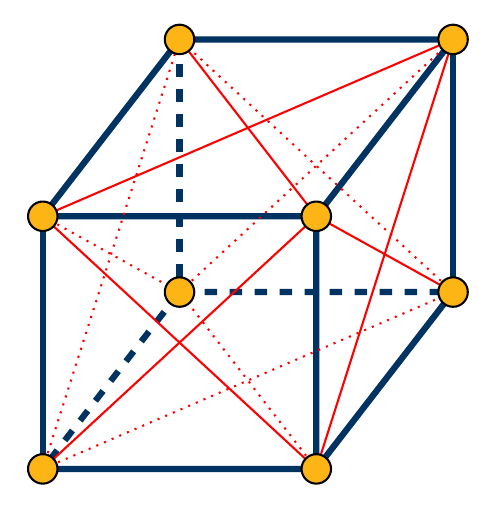}
\end{subfigure}
\caption{three-dimensional hypercube (left) and extended hypercube (right).}
\label{fig:hypercubes}
\end{figure}

\begin{exampleS}[Hypercube graph]  \label{exS:Hypercube}
Consider the $N$-dimensional hypercube, where each of $n=2^N$ vertices is involved in $N$ edges; 
see the left hand side of Figure \ref{fig:hypercubes}. 
This is an $N$-regular graph --- that is, $d_i=h_i = N$ for all $i$ --- with the total number of edges in the graph equaling $2^{N-1}$. Here,
$$ 
\lambda_2 = \frac{2}{N} = O((\ln n)^{-1}).
$$
Thus, $\lambda_2  \, h_i$ is constant in $n$. An application of Theorem \ref{thm:firstorderbound} yields 
\begin{align*}
1 + o(1)
 \,   \leq    \, \frac{N\, {\rm var}( \hat \alpha_i )}{\sigma^2}    \,     \leq   \,
 \frac{3}{2}  + o(1) .
\end{align*}
From this, we obtain the convergence rate result
$(\hat{\alpha}_i- \alpha_i)   = O_p\left( (\ln n)^{-1/2} \right)$. 
\end{exampleS}

Theorem \ref{thm:firstorderbound} allows to establish the convergence rate for the hypercube, but the conditions are too stringent to obtain \eqref{FirstOrderRate}.
The reason is that $h_i$ does not increase fast enough to ensure that $\lambda_2\,h_i\rightarrow \infty$.
The following example deals with an extended hypercube and illustrates that, despite $\lambda_2 \rightarrow 0$, we still have 
$\lambda_2\,h_i\rightarrow \infty$ in this case.

\begin{exampleS}[Extended Hypercube graph]
Start with the $N$-dimensional hypercube ${\mathcal G}$ from the previous example and add edges between all path-two neighbors in ${\mathcal G}$; see the right hand side of Figure \ref{fig:hypercubes} for an example. The resulting graph still has $n=2^N$ vertices, but now has $N(N+1) \, 2^{N-1}$  edges. Here,
$$
d_i = h_i = \frac{N(N+1)}{2}, \qquad \lambda_2 = \frac{4}{N+1},
$$
so that $\lambda_2\,h_i\rightarrow \infty$ holds, despite $\lambda_2 \rightarrow 0$
as $n \rightarrow \infty$.
Theorem~\ref{thm:firstorderbound} therefore implies \eqref{FirstOrderRate} in this example.
\end{exampleS}

The next example shows that our bound can still be informative if $h_i$ is finite.

\begin{figure}
\centering
\begin{subfigure}{.5\textwidth}
  \centering
  \includegraphics[width=.7\textwidth]{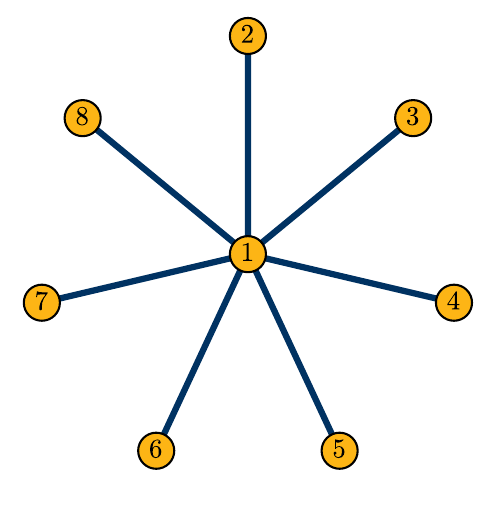}
\end{subfigure}%
\begin{subfigure}{.5\textwidth}
  \centering
  \includegraphics[width=.7\textwidth]{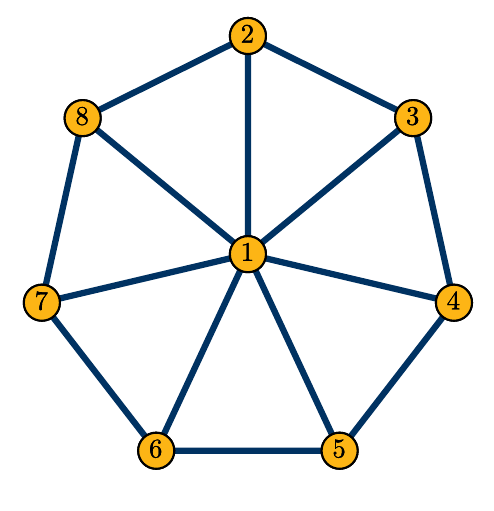}
\end{subfigure}
\caption{Star graph (left) and Wheel graph (right) for $n=8$.}
\label{fig:starwheel}
\end{figure}

\begin{exampleS}[Star graph]
\label{exS:Star}
Consider a Star graph around the central vertex $1$, that is, the graph with $n$ vertices and edges
$$
E= \lbrace (1,j): 2\leq j \leq n \rbrace;
$$
see the left hand side of Figure \ref{fig:starwheel}.
Here, $\lambda_2=1$ for any $n$ while $d_1=n-1$, $h_1=1$ and $d_i=1$, $h_i=n-1$ for $i\neq 1$. 
For $i=1$ one finds that the bounds in 
Theorem~\ref{thm:firstorderbound} imply that 
 $ {\rm var}( \hat \alpha_1  )  = O(n^{-1})$, and so
$$
(\hat{\alpha}_1- \alpha_1)   = O_p\left(  n^{-1/2} \right).
$$
In contrast, for $i \neq 1$ we find $\lambda_2\,h_i\rightarrow \infty$ and thus, although \eqref{FirstOrderRate} holds, these $\alpha_i$ cannot be estimated consistently as  $d_i = 1$. 
\end{exampleS}

The previous example also illustrates that $\lambda_2$ can be large despite having many vertices with small degrees. It is largely due to this property that we prefer 
to measure global connectivity by $\lambda_2$ and not by the  ``algebraic connectivity''
(the second smallest eigenvalue of ${\mathbold L}$; see, e.g., \citealt{Chung1997}), which has been studied more extensively.

Our last example shows the effect on the upper bound in Theorem \ref{thm:firstorderbound}
when neighbors themselves are more strongly connected.

\begin{exampleS}[Wheel graph]
\label{exS:Wheel}
The Wheel graph is obtained on combining a Star graph centered at vertex $1$ with a Cycle graph on the remaining $n-1$ vertices; see the right hand side of Figure \ref{fig:starwheel}. Thus, a Wheel graph contains strictly more edges than the underlying Star graph, although none of these involve the central vertex directly. From \cite{Butler2016}, we have
$$
\lambda_2 = \min \left\lbrace \frac{4}{3}, 1-\frac{2}{3}\cos \left(\frac{2\pi}{n}\right)\right\rbrace ,
$$
which satisfies $\lambda_2\geq1$ only for $n\leq 4$, and converges to $1/3$ at an exponential rate. However, while, as in the Star graph, $d_1=n-1$, we now have that $h_i=3$ for all $i\neq 1$. Hence, $\lambda_2\, h_1 > 1$ for any finite $n$ and the upper bound in Theorem \ref{thm:firstorderbound} is strictly smaller than in the Star graph.
\end{exampleS}

The last two examples also illustrate  that adding edges to the graph (in this case, to obtain the Wheel graph from
the Star graph) can result in a decrease of our measure of global connectivity~$\lambda_2$.
This is not a problem, however, for our results as we only require that $\lambda_2$ be sufficiently different from zero. The Wheel graph with $\lambda_2 \geq 1/3$, for example, clearly describes a very well globally connected graph by that measure.

\section{Variance bounds for differences}
\setcounter{equation}{0}

Our focus in the main text has been inference on the $\alpha_i$, under the constraint in \eqref{eq:constraint}, $\sum_i d_i \alpha_i = 0$. An alternative to normalizing the parameters that may be useful in certain applications is to focus directly on the differences $\alpha_i-\alpha_j$ for all $i\neq j$. An example where this is the case is \cite{FinkelsteinGentzkowWilliams2016}. We give a corresponding version of Theorem \ref{thm:firstorderbound} here.

Let $d_{ij}:= \sum_{k\in V} (\mathbold{A})_{ik} \, (\mathbold{A})_{jk}$. for an unweighted graph $d_{ij} = \left| [i] \cap [j] \right|$, the number of vertices that are neighbors of both $i$ and $j$. Write 
\begin{align*}
h_{ij} &:= \left\{
\begin{array}{cl} 
\displaystyle \left(\frac{1}{d_{ij}} \sum_{k\in V} \frac{(\mathbold{A})_{ik} \, (\mathbold{A})_{jk} }{d_k}\right)^{-1}
& \text{for $d_{ij} \neq 0$},
\\ 
\infty & \text{for $d_{ij} =0$},
\end{array}
\right.
\end{align*}
for the corresponding harmonic mean of the degrees of the vertices $ k\in  [i] \cap [j]$. 
We have the following theorem.

\begin{theoremS}[First-order bound for differences] \label{thm:firstorderbound2}
Let $\mathcal{G}$ be connected. Then
\begin{align*}
&\sigma^2  \left( \frac{1} {d_i} + \frac{1} {d_j} -  \frac{2 (\mathbold{A})_{ij}} {d_i d_j}   \right)
\\
& \qquad  \qquad    \qquad 
\leq 
{\rm var} (\hat{\alpha}_i-\hat{\alpha}_j)
\leq 
\sigma^2  \left( \frac{1} {d_i} + \frac{1} {d_j} -  \frac{2 (\mathbold{A})_{ij}} {d_i d_j}   \right)
+
\frac{\sigma^2} {\lambda_2} \left( \frac 1 {d_i h_i} + \frac 1 {d_j h_j} - \frac {2 \, d_{ij}} {d_i  d_j  h_{ij}} \right) .
\end{align*}
\end{theoremS}	

\bigskip

\noindent
For a simple graph $\mathcal{G}$, when $[i]=[j]$ but $i\notin [j]$ and $i\notin [j]$, that is, when vertices $i$ and $j$ share exactly the same neighbors and are not connected themselves, the theorem implies
\begin{align}
   \label{FirstOrderRateDiff}
{\rm var} (\hat{\alpha}_i-\hat{\alpha}_j)  =
\sigma^2  \left( \frac{1} {d_i} + \frac{1} {d_j}    \right),
\end{align}
as, in that case, both $(\mathbold{A})_{ij}$ and the second term in the upper bound in Theorem \ref{thm:firstorderbound2} are zero.

\section{Alternative normalization} \label{app:AlternNorm}

If we change the normalization constraint in the least-squares minimization problem \eqref{eq:ols} to 
$$
\sum_{i=1}^n \alpha_i = 0,
$$
we obtain the estimator $\hat{\mathbold{\alpha}}^{\, \diamond} = \mathbold{M}_{\mathbold{\iota}}  \hat{\mathbold{\alpha}}$,
where $ \mathbold{M}_{\mathbold{\iota}}  = \mathbold{I}_n - n^{-1} \mathbold{\iota}_n \mathbold{\iota}_n'$
is the projector orthogonal to $\mathbold{\iota}_n$.
We then have $  {\rm var}\left( \hat{\mathbold{\alpha}}^{\, \diamond} \right) = \sigma^2  \mathbold{L}^\dagger$,
because this variance needs to satisfy $  {\rm var}\left( \hat{\mathbold{\alpha}}^{\, \diamond} \right)  \mathbold{\iota}_n = 0$,
and the Moore-Penrose pseudoinverse  guarantees that the nullspace of  $\mathbold{L}$  equals the nullspace
of  $\mathbold{L}^\dagger$. Thus, changing the normalization corresponds to changing the particular pseudoinverse of $\mathbold{L}$
that features in the expression for the variance.
From $\hat{\mathbold{\alpha}}^{\, \diamond} = \mathbold{M}_{\mathbold{\iota}}  \hat{\mathbold{\alpha}}$ we
find
\begin{align*}
      {\rm var}\left( \hat{\mathbold{\alpha}}^{\, \diamond} \right) 
      &= \mathbold{M}_{\mathbold{\iota}}   {\rm var}\left( \hat{\mathbold{\alpha}}  \right)  \mathbold{M}_{\mathbold{\iota}} ,
\end{align*}
which thus also shows that $ \mathbold{L}^\dagger =  \mathbold{M}_{\mathbold{\iota}} \mathbold{L}^\pinv  \mathbold{M}_{\mathbold{\iota}}$.
We have $\mathbold{L}^\pinv \leq \lambda_2^{-1} \mathbold{D}^{-1}$, and therefore
$ \mathbold{L}^\dagger \leq \lambda_2^{-1}   \mathbold{M}_{\mathbold{\iota}}  \mathbold{D}^{-1}   \mathbold{M}_{\mathbold{\iota}} $.
We thus find
 ${\rm var}( \hat{\alpha}^{\, \diamond}_i ) = \sigma^2 \mathbold{e}_i' \mathbold{L}^\dagger \mathbold{e}_i
  \leq  \lambda_2^{-1}  \sigma^2  \mathbold{e}_i'   \mathbold{M}_{\mathbold{\iota}}  \mathbold{D}^{-1}   \mathbold{M}_{\mathbold{\iota}}  \mathbold{e}_i$,
and  evaluating the last expression gives the following theorem.

\begin{theoremS}[Global bound under alternative normalization] \label{thm:eigenvaluebound3}
Let $\mathcal{G}$ be connected. Then
$$
{\rm var}( \hat{\alpha}^{\, \diamond}_i )
\leq 
\frac{1}{d_i} \frac{\sigma^2}{\lambda_2} 
\left( 1   + \frac{d_i} {n \, h} \right) .$$
\end{theoremS}
\noindent
Notice that $d_i / (n \, h) \leq 1/h \leq 1$,
and therefore ${\rm var}( \hat{\alpha}^{\, \diamond}_i ) \leq \frac{2}{d_i} \frac{\sigma^2}{\lambda_2} $.
For the estimator $\hat \alpha_i$ obtained under the normalization in the main text
we immediately find from \eqref{eq:var1} and $\left( \mathbold{S}^\dagger \right)_{ii} \leq \lambda_2^{-1}$
that ${\rm var}( \hat{\alpha}_i ) \leq \frac{1}{d_i} \frac{\sigma^2}{\lambda_2} $.
Thus,  for sequences of growing networks
we find the pointwise consistency results
$(\hat{\alpha}^{\, \diamond}_i-\alpha_i) \overset{p}{\rightarrow} 0$
and $(\hat{\alpha}_i-\alpha_i) \overset{p}{\rightarrow} 0$
for both estimators, under the sufficient condition $\lambda_2 \, d_i \rightarrow \infty$.

Analogously one can extend Theorem~\ref{thm:firstorderbound} from $\hat \alpha_i $ to $\hat{\alpha}^{\, \diamond}_i $ as
follows.

\begin{theoremS}[First-order bound under alternative normalization] \label{thm:firstorderbound3}
Let $\mathcal{G}$ be connected. Then
\begin{align*}
 \frac{\sigma^2} {d_i} \left( 1 -  \frac 2 n \right)   - \frac {2 \, \sigma^2} {n \, h^{(2)}_i}  
 \,   \leq    \,  {\rm var}( \hat{\alpha}^{\, \diamond}_i  )  \,     \leq   \,
   \frac{\sigma^2} {d_i}   \left( 1  +  \frac{1}  {\lambda_2 h_i}  \right)  
   +  \frac{ \sigma^2} h  \left( \frac 2 {n}  + \frac{1} {\lambda_2 \, H}    \right)    ,
\end{align*}
where $h^{(2)}_i =  \left( \frac{1}{d_i} \sum_{j \in [i]} \frac{(\mathbold{A})_{ij}}{d_j} \right)^{-1}$,
and $h$ and $H$ defined in the main text.
\end{theoremS}
\noindent
Analogous to \eqref{FirstOrderRate} in the main text we thus find
\begin{align*}
    {\rm var}( \hat \alpha_i^{\, \diamond}  ) = \frac{\sigma^2}{d_i} + o(d^{-1}_i),
\end{align*}
provided that $\lambda_2 h_i \rightarrow \infty$ 
and $n h / d_i \rightarrow \infty $ and $n h_i^{(2)} / d_i \rightarrow \infty $
and $\lambda_2 \, h \, H / d_i \rightarrow \infty $
as $n\rightarrow\infty$. 
Therefore, under plausible assumptions on the sequence of growing networks we find the same asymptotic properties
for $ \hat \alpha_i^{\, \diamond} $ as for  $\hat \alpha_i $. The particular choice of normalization in the main text
is not necessary for our main results, but it makes all derivations as well as the presentation of the results more convenient.

\section{Proofs}
\label{sup:Proofs}

\paragraph{PROOF OF THEOREM~\ref{theorem:existence} (EXISTENCE)}
\addcontentsline{toc}{subsection}{~~Proof of Theorem~\ref{theorem:existence}}
\mbox{} \newline
The estimator is defined by the constraint minimization problem in \eqref{eq:ols}.
For convenience we express the constraint in quadratic form, 
$ \left( \mathbold{a}^\prime \mathbold{d}  \right)^2 = 0$.
By introducing the Lagrange multiplier $\lambda > 0$ we can write 
\begin{align*} 
\check{\mathbold{\alpha}}
&=
\arg 
\min_{ \mathbold{a}\in\mathbb{R}^n } \; \;
(\mathbold{y}-\mathbold{B}\mathbold{a})^\prime   \mathbold{M}_{\mathbold{X}}
(\mathbold{y}-\mathbold{B}\mathbold{a}) 
+ \lambda   \left( \mathbold{a}^\prime \mathbold{d}   \right)^2 .
\end{align*}
Solving the corresponding first-order condition we obtain
\begin{align}
   \check{\mathbold{\alpha}}  
   &= 
     \left( \mathbold{B}'  \mathbold{M}_{\mathbold{X}} \mathbold{B} + \lambda  \, \mathbold{d}  \, \mathbold{d}'  \right)^{-1}  \mathbold{B}'  \mathbold{M}_{\mathbold{X}} \mathbold{y} 
    \nonumber \\
   &= \mathbold{D}^{-1/2}  \left(  \mathbold{S}_{\mathbold{X}}  + \lambda  \, \mathbold{\psi}  \, \mathbold{\psi}'  \right)^{-1}  \mathbold{D}^{-1/2}  
   \mathbold{B}' \mathbold{y}  ,
   \label{AlgebraSx}
\end{align}
where $\mathbold{S}_{\mathbold{X}} := \mathbold{D}^{-1/2} \mathbold{B}'  \mathbold{M}_{\mathbold{X}} \mathbold{B} \mathbold{D}^{-1/2} $
and
$\mathbold{\psi} := \mathbold{D}^{1/2} \mathbold{\iota}_n  = \mathbold{D}^{-1/2}  \mathbold{d}$.
Since we assume that the graph is connected we have $d_i>0$ for all $i$, that is,
$ \mathbold{D}$ is invertible. 
Our assumption $\mathrm{rank}((\mathbold{X},\mathbold{B}))=p+n-1$ implies that
$\mathrm{rank}(\mathbold{B}'  \mathbold{M}_{\mathbold{X}} \mathbold{B})=n-1$, that is, 
the zero eigenvalue of $\mathbold{B}'  \mathbold{M}_{\mathbold{X}} \mathbold{B}$ has multiplicity one.
By construction of $\mathbold{B}$ we have $\mathbold{B} \mathbold{\iota}_n=0$, that is, 
the zero eigenvector of $\mathbold{B}'  \mathbold{M}_{\mathbold{X}} \mathbold{B}$ is given by
$ \mathbold{\iota}_n$. It follows that the zero eigenvalue $\mathbold{S}_{\mathbold{X}}$ has multiplicity one
and eigenvector  $ \mathbold{\psi} $.  
This explains why  the matrix $\mathbold{S}_{\mathbold{X}} + \lambda  \, \mathbold{\psi}  \, \mathbold{\psi}' $ is invertible, 
which we already used in \eqref{AlgebraSx}.
Furthermore, the matrices $  \mathbold{S}_{\mathbold{X}} $
and $ \mathbold{\psi}  \, \mathbold{\psi}'  $ commute, and by properties of the Moore-Penrose inverse
we thus have 
\begin{align}
\left(\mathbold{S}_{\mathbold{X}} + \lambda  \, \mathbold{\psi}  \, \mathbold{\psi}'  \right)^{-1} 
  =    \mathbold{S}_{\mathbold{X}}^\dagger + \lambda^{-1 }
  \left(\mathbold{\psi}  \, \mathbold{\psi}'  \right)^\dagger.
    \label{InverseS}
\end{align}
We furthermore have 
\begin{align} 
    \left(\mathbold{\psi}  \, \mathbold{\psi}'  \right)^\dagger =  m^{-2} \, \mathbold{\psi}  \, \mathbold{\psi}' ,
    \label{InversePsi}
\end{align}    
where $m = \mathbold{\psi}'  \mathbold{\psi}$ is the total number of observations.
Because $ \mathbold{B} \mathbold{\iota}_n = 0$, the contribution from 
$ \left( \mathbold{\psi}  \, \mathbold{\psi}'  \right)^\dagger$ drops out of \eqref{AlgebraSx}, and we obtain
\begin{align*}
  \check{\mathbold{\alpha}}  
  &= 
   \mathbold{D}^{-1/2}  \mathbold{S}_{\mathbold{X}}^\dagger  \mathbold{D}^{-1/2}  
   \mathbold{B}' \mathbold{y}
   =    \left( \mathbold{B}'  \mathbold{M}_{\mathbold{X}} \mathbold{B} \right)^\pinv  
   \mathbold{B}' \mathbold{y} ,
\end{align*}         
according to the definition of the pseudoinverse $^\pinv $ in the main text.
Notice that $\check{\mathbold{\alpha}}$ given in the last display does not depend on $\lambda$,
and automatically satisfies the constraint $\mathbold{d}'\check{\mathbold{\alpha}}=0$,
that is, any value of $\lambda$ can be chosen in the above derivation.
  \qed

\paragraph{PROOF OF THEOREMS~\ref{thm:firstorderbound} AND~\ref{thm:firstorderbound2} (VARIANCE BOUNDS)}
\addcontentsline{toc}{subsection}{~~Proof of Theorems~\ref{thm:firstorderbound} and \ref{thm:firstorderbound2}}
\mbox{} \newline
We first show that, if $\mathcal{G}$ is connected, then
    \begin{align}
         0  \leq  
         \left[  {\rm var}( \hat {\mathbold \alpha}  )  - 
        \sigma^2   \left(  \mathbold{D}^{-1}  
        +   \mathbold{D}^{-1} \mathbold{A}  \mathbold{D}^{-1}
         -  2 \, m^{-1} \,   \mathbold{\iota}_n \mathbold{\iota}_n'     \right) \right] 
       \leq   \frac{\sigma^2} {\lambda_2}  
          \mathbold{D}^{-1}  \mathbold{A}  \mathbold{D}^{-1}    \mathbold{A}  \mathbold{D}^{-1} .
        \label{BoundFirstOderMatrix}  
    \end{align}
Theorems~\ref{thm:firstorderbound} and~\ref{thm:firstorderbound2} will then follow readily. 
Analogous to \eqref{InverseS} we also hav
$\left(\mathbold{S} + \lambda  \, \mathbold{\psi}  \, \mathbold{\psi}'  \right)^{-1} 
  =    \mathbold{S}^\dagger + \lambda^{-1 }
  \left(\mathbold{\psi}  \, \mathbold{\psi}'  \right)^\dagger$.
Using this and     \eqref{InversePsi} we find
\begin{align*}
    \mathbold{I}_n
      &=    \left(\mathbold{S} + \lambda  \, \mathbold{\psi}  \, \mathbold{\psi}'  \right)^{-1} 
       \left(\mathbold{S} + \lambda  \, \mathbold{\psi}  \, \mathbold{\psi}'  \right)   
      \\
      &=  \left( \mathbold{S}^\dagger + \lambda^{-1 } \, m^{-2} \, \mathbold{\psi}  \, \mathbold{\psi}'  \right)
           \left(\mathbold{S} + \lambda  \, \mathbold{\psi}  \, \mathbold{\psi}'  \right) ,
\end{align*}
and since $\mathbold{S}  \mathbold{\psi}  = 0$ and $ \mathbold{\psi}'  \mathbold{\psi} = m$ we thus find that
$
      \mathbold{S}^\dagger \mathbold{S}  =     \mathbold{I}_n - m^{-1}  \mathbold{\psi}   \mathbold{\psi} ' ,
$
which is simply the idempotent matrix that projects orthogonally to $\mathbold{\psi}$.
We thus find
$
      \mathbold{L}^\pinv  \mathbold{L} 
            =   \mathbold{D}^{-1/2 }   \mathbold{S}^\dagger     \mathbold{S}   \mathbold{D}^{1/2 } 
            =  \mathbold{I}_n - m^{-1}    \mathbold{\iota}_n    \mathbold{d}'   .
$
Plugging in
   $\mathbold{L} = \mathbold{D} - \mathbold{A}$, and then solving for $\mathbold{L}^\pinv$ gives 
    \begin{align} 
        \mathbold{L}^\pinv &=   \mathbold{D}^{-1}   + \mathbold{L}^\pinv  \mathbold{A}    \mathbold{D}^{-1}  
         - m^{-1} \, \mathbold{\iota}_n \mathbold{\iota}_n'    .
       \label{IntermediateL1}  
 \end{align}
The Laplacian is symmetric, and so transposition gives
  \begin{align}
        \mathbold{L}^\pinv &=   \mathbold{D}^{-1}   +   \mathbold{D}^{-1}   \mathbold{A}   \mathbold{L}^\pinv 
         - m^{-1} \,   \mathbold{\iota}_n \mathbold{\iota}_n'   .
       \label{IntermediateL2}  
   \end{align}
Replacing $\mathbold{L}^\pinv$
on the right-hand side of \eqref{IntermediateL1} by the expression for $ \mathbold{L}^\pinv$ given by \eqref{IntermediateL2},
and also using that $ \mathbold{D}^{-1}   \mathbold{A}   \mathbold{\iota}_n =   \mathbold{\iota}_n$,
yields 
    \begin{align}
  \mathbold{L}^\pinv  
  &=       \mathbold{D}^{-1}  
        +   \mathbold{D}^{-1} \mathbold{A}  \mathbold{D}^{-1} 
       +    \mathbold{D}^{-1}  \mathbold{A}   \mathbold{L}^\pinv   \mathbold{A}  \mathbold{D}^{-1}
       -  2 m^{-1} \,   \mathbold{\iota}_n \mathbold{\iota}_n'    .
        \label{IntermediateL3}     
\end{align}
Re-arranging this equation allows us to write
 \begin{align*}
  \mathbold{L}^\pinv  
  - \left(  \mathbold{D}^{-1}  
        +   \mathbold{D}^{-1} \mathbold{A}  \mathbold{D}^{-1} 
       -   2 \, m^{-1} \,   \mathbold{\iota}_n \mathbold{\iota}_n'    \right)
      &=   \mathbold{D}^{-1}  \mathbold{A}   \mathbold{L}^\pinv   \mathbold{A}  \mathbold{D}^{-1}      .
\end{align*}
From $ \mathbold{L}^* = \mathbold{D}^{-1/2}  \mathbold{S}^\dagger  \mathbold{D}^{-1/2} $
and $\mathbf{0} \leq  \mathbold{S}^\dagger   \leq \lambda_2^{-1}     \mathbold{I}_n$
we obtain $\mathbf{0} \leq \mathbold{L}^\pinv \leq \lambda_2^{-1}    \mathbold{D}^{-1}$,
and therefore 
$$
\mathbf{0} \leq  \mathbold{D}^{-1}  \mathbold{A}   \mathbold{L}^\pinv   \mathbold{A}  \mathbold{D}^{-1} 
 \leq    \lambda_2^{-1}    \mathbold{D}^{-1}  \mathbold{A}  \mathbold{D}^{-1}   \mathbold{A}  \mathbold{D}^{-1}.
$$
Put together this yields 
    \begin{align*}
   0 \leq \mathbold{L}^\pinv  
  - \left(   \mathbold{D}^{-1}  
        +   \mathbold{D}^{-1} \mathbold{A}  \mathbold{D}^{-1} 
       -  2 \, m^{-1} \,   \mathbold{\iota}_n \mathbold{\iota}_n'     
     \right)
     &\leq     \lambda_2^{-1}    \mathbold{D}^{-1}  \mathbold{A}  \mathbold{D}^{-1}   \mathbold{A}  \mathbold{D}^{-1}    ,
\end{align*}
and multiplication with $\sigma^2 $ gives the bounds stated in \eqref{BoundFirstOderMatrix}.

To show Theorems~\ref{thm:firstorderbound} and~\ref{thm:firstorderbound2} we calculate, for $i\neq j$,
\begin{equation*}
\begin{aligned}[c]
     {\mathbold e}_i' \,  \mathbold{D}^{-1} \,      {\mathbold e}_i
     &= d_i^{-1}, \\
     {\mathbold e}_i' \,  \mathbold{D}^{-1} \,      {\mathbold e}_j
     &= 0,\\
     {\mathbold e}_i'  \, \mathbold{D}^{-1} \mathbold{A}  \mathbold{D}^{-1}  \,  {\mathbold e}_i
     &=0 , \\
     {\mathbold e}_i'  \, \mathbold{D}^{-1} \mathbold{A}  \mathbold{D}^{-1}  \,  {\mathbold e}_j
     &= d_i^{-1} d_j^{-1}  ({\mathbold A})_{ij}	 ,
\end{aligned}
\qquad\qquad
\begin{aligned}[c]
     {\mathbold e}_i'  \, \mathbold{D}^{-1} \mathbold{A}  \mathbold{D}^{-1}   \mathbold{A}  \mathbold{D}^{-1} \,  {\mathbold e}_i
     &=  d_i^{-1}  h_i^{-1} ,
   \\  
     {\mathbold e}_i'  \, \mathbold{D}^{-1} \mathbold{A}  \mathbold{D}^{-1}   \mathbold{A}  \mathbold{D}^{-1} \,  {\mathbold e}_j
     &= d_i^{-1} d_j^{-1}  d_{ij}  h_{ij}^{-1} ,
     \\
     {\mathbold e}_i' \,   \mathbold{\iota}_n \mathbold{\iota}'_n       {\mathbold e}_i
     &= 1,
     \\
     {\mathbold e}_i' \,   \mathbold{\iota}_n \mathbold{\iota}'_n \,    {\mathbold e}_j
     &= 1,
\end{aligned}
\end{equation*}
where $\mathbold{e}_i$ is the vector that has one as its $i^{\mathrm{th}}$ entry and zeros elsewhere.
Combining these results with \eqref{BoundFirstOderMatrix}
gives the bounds on, respectively, 
  ${\rm var}( \hat \alpha_i  )   =  {\mathbold e}_i'  {\rm var}( \hat {\mathbold \alpha}  )    {\mathbold e}_i$ and  ${\rm var}( \hat \alpha_i - \hat \alpha_j    )   =  ( {\mathbold e}_i - {\mathbold e}_j )'  {\rm var}( \hat {\mathbold \alpha}  )    ( {\mathbold e}_i - {\mathbold e}_j )$ stated in the theorems. \qed

\paragraph{PROOF OF THEOREMS~\ref{thm:eigenvaluebound3} AND \ref{thm:firstorderbound3}}
\addcontentsline{toc}{subsection}{~~Proof of Theorems~\ref{thm:eigenvaluebound3} and \ref{thm:firstorderbound3}}
\mbox{} \newline
     Using that $ \mathbold{L}^*  \leq \lambda_2^{-1}\mathbold{D}^{-1}$ we find that
     \begin{align*}
           {\rm var}( \hat{\alpha}^{\, \diamond}_i )
           &=
          \mathbold{e}_i'  {\rm var}\left( \hat{\mathbold{\alpha}}^{\, \diamond} \right)   \mathbold{e}_i
      =\mathbold{e}_i'   \mathbold{M}_{\mathbold{\iota}}   {\rm var}\left( \hat{\mathbold{\alpha}}  \right)  \mathbold{M}_{\mathbold{\iota}}
       \mathbold{e}_i
        = \sigma^2 \mathbold{e}_i' \mathbold{M}_{\mathbold{\iota}} \mathbold{L}^* \mathbold{M}_{\mathbold{\iota}} \mathbold{e}_i
        \\
         &\leq  \lambda_2^{-1}  \sigma^2  \mathbold{e}_i'   \mathbold{M}_{\mathbold{\iota}}  \mathbold{D}^{-1}   \mathbold{M}_{\mathbold{\iota}}  \mathbold{e}_i ,
     \end{align*}
     and we calculate 
     \begin{align}
          \mathbold{e}_i'   \mathbold{M}_{\mathbold{\iota}}  \mathbold{D}^{-1}   \mathbold{M}_{\mathbold{\iota}}  \mathbold{e}_i
          &=   \mathbold{e}_i'  \mathbold{D}^{-1}   \mathbold{e}_i
             - \frac 2 n      \mathbold{e}_i'  \mathbold{D}^{-1}     \mathbold{\iota}_n
             + \frac 1 {n^2 }   \mathbold{\iota}_n' \mathbold{D}^{-1}     \mathbold{\iota}_n
       \nonumber \\
         &=
          \frac 1 {d_i} - \frac{2} {n \, d_i} + \frac 1 {n \, h} .
          \label{BoundEMDME}
     \end{align}
     Combing those results gives the statement of Theorem~\ref{thm:eigenvaluebound3}

\bigskip     

   Next, multiplying $\mathbold{M}_{\mathbold{\iota}}$ from the left and right to the matrix bounds \eqref{BoundFirstOderMatrix}
   and using    ${\rm var}\left( \hat{\mathbold{\alpha}}^{\, \diamond} \right) 
      = \mathbold{M}_{\mathbold{\iota}}   {\rm var}\left( \hat{\mathbold{\alpha}}  \right)  \mathbold{M}_{\mathbold{\iota}}$
   gives
    \begin{align*}
         0  \leq  
         \left[  {\rm var}\left( \hat{\mathbold{\alpha}}^{\, \diamond} \right)   - 
        \sigma^2   \mathbold{M}_{\mathbold{\iota}}   \left(  \mathbold{D}^{-1}  
        +     \mathbold{D}^{-1} \mathbold{A}  \mathbold{D}^{-1}  
           \right)\mathbold{M}_{\mathbold{\iota}}    \right] 
       \leq   \frac{\sigma^2} {\lambda_2}  
         \mathbold{M}_{\mathbold{\iota}}   \mathbold{D}^{-1}  \mathbold{A}  \mathbold{D}^{-1}    \mathbold{A}  \mathbold{D}^{-1} \mathbold{M}_{\mathbold{\iota}}  ,
    \end{align*}
    and therefore
    \begin{align*}
         0  \leq  
         \left[    {\rm var}( \hat{\alpha}^{\, \diamond}_i )   - 
       \sigma^2   \mathbold{e}_i'  \mathbold{M}_{\mathbold{\iota}}   \left(  \mathbold{D}^{-1}  
        +     \mathbold{D}^{-1} \mathbold{A}  \mathbold{D}^{-1}  
           \right)\mathbold{M}_{\mathbold{\iota}}   \mathbold{e}_i  \right] 
       \leq   \frac{\sigma^2} {\lambda_2}  
         \mathbold{e}_i'  \mathbold{M}_{\mathbold{\iota}}   \mathbold{D}^{-1}  \mathbold{A}  \mathbold{D}^{-1}    \mathbold{A}  \mathbold{D}^{-1} \mathbold{M}_{\mathbold{\iota}}   \mathbold{e}_i  .
    \end{align*}
    We already calculated 
    $   \mathbold{e}_i'   \mathbold{M}_{\mathbold{\iota}}  \mathbold{D}^{-1}   \mathbold{M}_{\mathbold{\iota}}  \mathbold{e}_i$ 
    in \eqref{BoundEMDME} above.
    We furthermore have 
    \begin{align*}
          \mathbold{e}_i'   \mathbold{M}_{\mathbold{\iota}}   \mathbold{D}^{-1} \mathbold{A}  \mathbold{D}^{-1}    \mathbold{M}_{\mathbold{\iota}}  \mathbold{e}_i
           &=  \mathbold{e}_i'  \mathbold{D}^{-1} \mathbold{A}  \mathbold{D}^{-1}   \mathbold{e}_i
             - \frac 2 n      \mathbold{e}_i'   \mathbold{D}^{-1} \mathbold{A}  \mathbold{D}^{-1}  \mathbold{\iota}_n
             + \frac 1 {n^2 } \,  \mathbold{\iota}_n' \ \mathbold{D}^{-1} \mathbold{A}  \mathbold{D}^{-1}   \mathbold{\iota}_n
             \\
           &=  0 
              - \frac 2 {n \, d_i} \sum_{j \in [i]}    \frac {( \mathbold{A}  )_{ij} } {d_j} 
               +  \frac 1 {n^2 } \, \sum_{j,k=1}^n    \frac {( \mathbold{A}  )_{jk} } {d_j d_k} ,
    \end{align*}
    and by applying  the Cauchy-Schwarz inequality  
    we find $\sum_{j,k}    \frac {( \mathbold{A}  )_{jk} } {d_j d_k}  \leq 
    \sum_{j,k}    \frac {( \mathbold{A}  )_{jk} } {d_j^2}  = \sum_j \frac 1 {d_j}$,
    and
    therefore
    \begin{align*}
            - \frac 2 {n \, h^{(2)}_i} 
            &\leq  \mathbold{e}_i'   \mathbold{M}_{\mathbold{\iota}}   \mathbold{D}^{-1} \mathbold{A}  \mathbold{D}^{-1}    \mathbold{M}_{\mathbold{\iota}}  \mathbold{e}_i
           \leq    \frac 1 {n \, h} .
    \end{align*}
    Similarly, $ \mathbold{e}_i'  \mathbold{M}_{\mathbold{\iota}}   \mathbold{D}^{-1}  \mathbold{A}  \mathbold{D}^{-1}    \mathbold{A}  \mathbold{D}^{-1} \mathbold{M}_{\mathbold{\iota}}   \mathbold{e}_i \geq 0$
    contains three terms, for which we have
    \begin{align*}
         \mathbold{e}_i'      \mathbold{D}^{-1}  \mathbold{A}  \mathbold{D}^{-1}    \mathbold{A}  \mathbold{D}^{-1}   \mathbold{e}_i
         &= \frac 1 {d_i \, h_i} ,
    \end{align*}
    \begin{align*}
         - \frac 2 n \, \mathbold{e}_i'      \mathbold{D}^{-1}  \mathbold{A}  \mathbold{D}^{-1}    \mathbold{A}  \mathbold{D}^{-1}   \mathbold{\iota}_n
         &= - \frac 2 {n \,d_i} \sum_{j \in [i]} \frac {( \mathbold{A}  )_{ij}} {d_j}  \sum_{k \in [j]} \frac {( \mathbold{A}  )_{jk}} {d_k}
         \leq 0   ,
    \end{align*}
    \begin{align*}
         \frac 1 {n^2} \,  \mathbold{\iota}_n'      \mathbold{D}^{-1}  \mathbold{A}  \mathbold{D}^{-1}    \mathbold{A}  \mathbold{D}^{-1}   \mathbold{\iota}_n
         &= \frac 1 {n^2} 
         \sum_{i,j,k}  \frac {( \mathbold{A}  )_{ij} ( \mathbold{A}  )_{jk}} {d_i d_j d_k}  
         \leq  \frac 1 {n^2} 
         \sum_{i,j,k}  \frac {( \mathbold{A}  )_{ij}^2  } {d_i^2 d_j}  
         = \frac 1 n   \sum_{i} \frac 1 {d_i h_i}  = \frac 1 {h \, H}
         ,
    \end{align*}
where in the last line we again applied     the Cauchy-Schwarz inequality,
and the definitions of the harmonic means $h$ and $H$ in the main text.
Combining the above gives the statement of Theorem~\ref{thm:firstorderbound3}.

\paragraph{PROOF OF THEOREM~\ref{thm:regressors} (COVARIATES)}
\addcontentsline{toc}{subsection}{~~Proof of Theorem~\ref{thm:regressors}}
\mbox{} \newline
Define the $n \times n$ matrix
\begin{align*}
\mathbold{C} := 
 \left(\mathbold{B}'  \mathbold{B}  \right)^\pinv
\mathbold{B}' {\mathbold X} \left(   {\mathbold X}' {\mathbold X} \right)^{-1}  {\mathbold X}' \mathbold{B} .
\end{align*}
Let $\lambda_i(\mathbold{C})$ denote the $i$th eigenvalue of $\mathbold{C}$, arranged in ascending order. $\mathbold{C}$ is similar to the positive semi-definite matrix
$$
\left(   {\mathbold X}' {\mathbold X} \right)^{-1/2}  {\mathbold X}' \mathbold{B}
 \left(\mathbold{B}'  \mathbold{B}  \right)^\pinv \mathbold{B}' {\mathbold X} \left(   {\mathbold X}' {\mathbold X} \right)^{-1/2},
$$
and since similar matrices share the same eigenvalues
we have $\lambda_1(\mathbold{C}) \geq 0$.  $\mathbold{C}$ is also similar to the matrix
$$
\mathbold{B} \left(\mathbold{B}'  \mathbold{B}  \right)^\pinv
\mathbold{B}' {\mathbold X} \left(   {\mathbold X}' {\mathbold X} \right)^{-1}  {\mathbold X}',
$$
which is the product of two projection matrices, whose spectral norm is thus bounded by one. Hence, $\lambda_n(\mathbold{C})\leq 1$. In addition, we must have $\lambda_i( \mathbold{C} ) \neq 1$ for any $1<i<n$ because, otherwise, 
${\rm rank}\left(   {\mathbold I}_n - \mathbold{C}  \right) < n$, which implies
that ${\rm rank}( \mathbold{B}'  {\mathbold M}_{\mathbold X} \mathbold{B})  < n-1$,
contradicting our non-collinearity assumption
(since the graph is connected we have ${\rm rank}( \mathbold{B}'\mathbold{B} ) = n-1 $,
which together with the non-collinearity assumption $\mathrm{rank}((\mathbold{X},\mathbold{B}))=p+n-1$
implies that ${\rm rank}( \mathbold{B}'  {\mathbold M}_{\mathbold X} \mathbold{B})  = n-1$).
 We therefore have $\lVert \mathbold{C}\rVert_2 < 1$,
implying that $ {\mathbold I}_m - \mathbold{C}$ is invertible.

Using \eqref{InverseS} and \eqref{InversePsi} with $\lambda=m^{-1}$ we find that
$
      \left(  \mathbold{B}'  {\mathbold M}_{\mathbold X} \mathbold{B}
    + m^{-1}  \mathbold{D}  {\mathbold \iota}_n  {\mathbold \iota}_n'  \mathbold{D}
    \right)^{-1} 
    = 
    (\mathbold{B}'  {\mathbold M}_{\mathbold X} \mathbold{B})^\pinv
   + m^{-1} \,  {\mathbold \iota}_n  {\mathbold \iota}_n' ,
$
or equivalently
\begin{align*}
      \mathbold{B}'  {\mathbold M}_{\mathbold X} \mathbold{B}
    + m^{-1}  \mathbold{D}  {\mathbold \iota}_n  {\mathbold \iota}_n'  \mathbold{D}
   &= 
   \left[  (\mathbold{B}'  {\mathbold M}_{\mathbold X} \mathbold{B})^\pinv
   + m^{-1} \,  {\mathbold \iota}_n  {\mathbold \iota}_n' \right]^{-1},
\end{align*}
and analogously we have
\begin{align}
      \mathbold{B}'    \mathbold{B}
    + m^{-1}  \mathbold{D}  {\mathbold \iota}_n  {\mathbold \iota}_n'  \mathbold{D}
   &= 
   \left[  (\mathbold{B}'    \mathbold{B})^\pinv
   + m^{-1} \,  {\mathbold \iota}_n  {\mathbold \iota}_n' \right]^{-1}.
  \label{PseudoinverseExpression} 
\end{align}
Subtracting the expressions in the last two displays gives
\begin{align*}
    \mathbold{B}' {\mathbold X} \left(   {\mathbold X}' {\mathbold X} \right)^{-1}  {\mathbold X}' \mathbold{B}
    &=   \left[  (\mathbold{B}'    \mathbold{B})^\pinv
   + m^{-1} \,  {\mathbold \iota}_n  {\mathbold \iota}_n' \right]^{-1}
   -  \left[  (\mathbold{B}'  {\mathbold M}_{\mathbold X} \mathbold{B})^\pinv
   + m^{-1} \,  {\mathbold \iota}_n  {\mathbold \iota}_n' \right]^{-1} ,
\end{align*}
and by multiplying with $ \left[  (\mathbold{B}'    \mathbold{B})^\pinv
   + m^{-1} \,  {\mathbold \iota}_n  {\mathbold \iota}_n' \right]$ from the left
   and $ \left[  (\mathbold{B}'  {\mathbold M}_{\mathbold X} \mathbold{B})^\pinv
   + m^{-1} \,  {\mathbold \iota}_n  {\mathbold \iota}_n' \right]$ from the right, and using $ \mathbold{B}  {\mathbold \iota}_n = 0 $, we obtain
\begin{align*}
 (\mathbold{B}'    \mathbold{B})^\pinv    \mathbold{B}' {\mathbold X} \left(   {\mathbold X}' {\mathbold X} \right)^{-1}  {\mathbold X}' \mathbold{B}
  (\mathbold{B}'  {\mathbold M}_{\mathbold X} \mathbold{B})^\pinv
    &=    (\mathbold{B}'  {\mathbold M}_{\mathbold X} \mathbold{B})^\pinv 
   -    (\mathbold{B}'    \mathbold{B})^\pinv ,
\end{align*}
which can equivalently be expressed as
$      \left(  {\mathbold I}_m - \mathbold{C} \right) (\mathbold{B}'  {\mathbold M}_{\mathbold X} \mathbold{B})^\pinv
      =       
  \left(\mathbold{B}'  \mathbold{B}    \right)^\pinv$.
We have already argued that   $\left(  {\mathbold I}_m - \mathbold{C} \right) $ is invertible, and therefore
\begin{align*}
     (\mathbold{B}'  {\mathbold M}_{\mathbold X} \mathbold{B})^\pinv
     &=         \left(  {\mathbold I}_m - \mathbold{C} \right)^{-1}
  \left(\mathbold{B}'  \mathbold{B}    \right)^\pinv .
\end{align*}
Since $\left\| \mathbold{C}  \right\|_2 < 1$ we can expand
 $   \left(  {\mathbold I}_m - \mathbold{C} \right)^{-1}$ in powers of $ \mathbold{C} $, as
\begin{align}
     (\mathbold{B}'  {\mathbold M}_{\mathbold X} \mathbold{B})^\pinv
     &= \sum_{r=0}^{\infty}    \mathbold{C}^r    \left(\mathbold{B}'  \mathbold{B}    \right)^\pinv .
     \label{PowerExpansionC}
\end{align}
Defining the $p \times p$ matrix
\begin{align*}
   \widetilde {\mathbold C} := 
\left(   {\mathbold X}' {\mathbold X} \right)^{-1/2}  {\mathbold X}' \mathbold{B} \left(\mathbold{B}'  \mathbold{B}  \right)^\pinv
\mathbold{B}' {\mathbold X} \left(   {\mathbold X}' {\mathbold X} \right)^{-1/2}  
\end{align*}
we can rewrite \eqref{PowerExpansionC}   as
\begin{align*}
     (\mathbold{B}'  {\mathbold M}_{\mathbold X} \mathbold{B})^\pinv
     &=  \left(\mathbold{B}'  \mathbold{B}    \right)^\pinv
     +  \left(\mathbold{B}'  \mathbold{B}  \right)^\pinv
\mathbold{B}' {\mathbold X} \left(   {\mathbold X}' {\mathbold X} \right)^{-1/2} 
   \left(  \sum_{r=0}^{\infty}  \widetilde {\mathbold C}^{\,r}   \right)
 \left(   {\mathbold X}' {\mathbold X} \right)^{-1/2}   {\mathbold X}' \mathbold{B}
  \left(\mathbold{B}'  \mathbold{B}    \right)^\pinv .
\end{align*}
The parameter $\rho$ defined in the main text satisfies
\begin{align*}
   \rho 
 &=  \left\lVert (\mathbold{X}^\prime \mathbold{X})^{-1/2}  \mathbold{X}^\prime \mathbold{M}_{\mathbold{B}} \mathbold{X}   (\mathbold{X}^\prime \mathbold{X})^{-1/2}  \right\rVert_2
 =   \left\lVert   {\mathbold I}_p -    \widetilde {\mathbold C}  \right\rVert_2
    = 1 - \lVert \widetilde {\mathbold C} \rVert_2,
\end{align*}   
that is,  we have $\lVert \widetilde {\mathbold C} \rVert_2 = 1 - \rho$, and since $ \widetilde {\mathbold C}$ is symmetric
and semi-definite this can equivalently be written as $ \widetilde {\mathbold C} \leq (1 - \rho) {\mathbold I}_p$.
Therefore, 
\begin{align*}
    \sum_{r=0}^{\infty}  \widetilde {\mathbold C}^{\,r}  
    \leq   \sum_{r=0}^{\infty} \left( 1 - \rho \right)^r \;  {\mathbold I}_p 
    =  \rho^{-1} \; {\mathbold I}_p .
\end{align*}
We thus have
\begin{align}
  (\mathbold{B}'  {\mathbold M}_{\mathbold X} \mathbold{B})^\pinv 
              -  \left(\mathbold{B}'  \mathbold{B}    \right)^\pinv  
       &=       \left(\mathbold{B}'  \mathbold{B}  \right)^\pinv
\mathbold{B}' {\mathbold X} \left(   {\mathbold X}' {\mathbold X} \right)^{-1/2} 
   \left(  \sum_{r=0}^{\infty}  \widetilde {\mathbold C}^{\,r}   \right)
 \left(   {\mathbold X}' {\mathbold X} \right)^{-1/2}   {\mathbold X}' \mathbold{B}
  \left(\mathbold{B}'  \mathbold{B}    \right)^\pinv  
\nonumber \\
    &\leq  \frac{1} {\rho}       \left(\mathbold{B}'  \mathbold{B}  \right)^\pinv
\mathbold{B}' {\mathbold X}  
 \left(   {\mathbold X}' {\mathbold X} \right)^{-1}   {\mathbold X}' \mathbold{B}
  \left(\mathbold{B}'  \mathbold{B}    \right)^\pinv    ,
   \label{BoundBMB}
\end{align}
and, therefore,
\begin{align*}
     {\rm var}\left(       \check\alpha_i \right)
            -  {\rm var}\left(       \hat \alpha_i \right)
        &= \sigma^2 \,  {\mathbold e}_i'   
          \left[    (\mathbold{B}'  {\mathbold M}_{\mathbold X} \mathbold{B})^\pinv 
              -  \left(\mathbold{B}'  \mathbold{B}    \right)^\pinv \right]  {\mathbold e}_i
 \\
    & \leq  
     \frac{ \sigma^2} {\rho} \,  {\mathbold e}_i'   
          \left[    \left(\mathbold{B}'  \mathbold{B}  \right)^\pinv
\mathbold{B}' {\mathbold X}  
 \left(   {\mathbold X}' {\mathbold X} \right)^{-1}   {\mathbold X}' \mathbold{B}
  \left(\mathbold{B}'  \mathbold{B}    \right)^\pinv \right]  {\mathbold e}_i  .
\end{align*}
Using the expression  \eqref{IntermediateL1} and \eqref{IntermediateL2}
for $ \left(\mathbold{B}'  \mathbold{B}    \right)^\pinv =  \mathbold{L}^\pinv $
 we obtain   
\begin{align*}
  &  {\mathbold e}_i'  \left(\mathbold{B}'  \mathbold{B}  \right)^\pinv
\mathbold{B}' {\mathbold X} \left(   {\mathbold X}' {\mathbold X} \right)^{-1}   {\mathbold X}' \mathbold{B}
  \left(\mathbold{B}'  \mathbold{B}    \right)^\pinv   {\mathbold e}_i
 \\ 
  &=     {\mathbold e}_i'  \mathbold{L}^\pinv
\mathbold{B}' {\mathbold X} \left(   {\mathbold X}' {\mathbold X} \right)^{-1}   {\mathbold X}' \mathbold{B}
 \mathbold{L}^\pinv  {\mathbold e}_i
\\
  &=     {\mathbold e}_i'  
  \left(   \mathbold{D}^{-1}   +   \mathbold{D}^{-1}   \mathbold{A}   \mathbold{L}^\pinv  \right)
\mathbold{B}' {\mathbold X} \left(   {\mathbold X}' {\mathbold X} \right)^{-1}   {\mathbold X}' \mathbold{B}
 \left(   \mathbold{D}^{-1}   + \mathbold{L}^\pinv  \mathbold{A}    \mathbold{D}^{-1}   \right) {\mathbold e}_i 
\\
  &\leq   T^{(1)}_i +  T^{(2)}_i + 2 \sqrt{T^{(1)}_i \, T^{(2)}_i}  ,
\end{align*}   
where
\begin{align*}
    T^{(1)}_i &:=  {\mathbold e}_i'   \mathbold{D}^{-1} 
                \mathbold{B}' {\mathbold X} \left(   {\mathbold X}' {\mathbold X} \right)^{-1}   {\mathbold X}' \mathbold{B}
                 \mathbold{D}^{-1}{\mathbold e}_i ,
         \\
   T^{(2)}_i &:=         {\mathbold e}_i'  
    \mathbold{D}^{-1}   \mathbold{A}   \mathbold{L}^\pinv 
\mathbold{B}' {\mathbold X} \left(   {\mathbold X}' {\mathbold X} \right)^{-1}   {\mathbold X}' \mathbold{B}
  \mathbold{L}^\pinv  \mathbold{A}    \mathbold{D}^{-1}    {\mathbold e}_i      ,   
\end{align*}
 and we used the Cauchy-Schwarz inequality to bound the mixed term.
Again, because similar matrices have the same eigenvalues we have
\begin{align*}
    \lVert
     \left( \mathbold{L}^\pinv \right)^{1/2}
\mathbold{B}' {\mathbold X} \left(   {\mathbold X}' {\mathbold X} \right)^{-1}   {\mathbold X}' \mathbold{B}
 \left( \mathbold{L}^\pinv \right)^{1/2} 
 \rVert_2
   &=  \lVert  \widetilde {\mathbold C} \rVert_2  = 1-\rho ,
\end{align*}
and therefore,
\begin{align*}
     T^{(2)}_i &=         {\mathbold e}_i'  
    \mathbold{D}^{-1}   \mathbold{A}  \left( \mathbold{L}^\pinv \right)^{1/2}
    \left[   \left( \mathbold{L}^\pinv \right)^{1/2}
\mathbold{B}' {\mathbold X} \left(   {\mathbold X}' {\mathbold X} \right)^{-1}   {\mathbold X}' \mathbold{B}
 \left( \mathbold{L}^\pinv \right)^{1/2} 
 \right]
   \left( \mathbold{L}^\pinv \right)^{1/2}  \mathbold{A}    \mathbold{D}^{-1}    {\mathbold e}_i 
   \\
    &\leq  (1-\rho) \; {\mathbold e}_i'  
    \mathbold{D}^{-1}   \mathbold{A}  \mathbold{L}^\pinv \mathbold{A}    \mathbold{D}^{-1}    {\mathbold e}_i 
   \\
    & \leq  
    \frac {1-\rho} {\lambda_2}
     {\mathbold e}_i'  
    \mathbold{D}^{-1}   \mathbold{A}  \mathbold{D}^{-1} \mathbold{A}    \mathbold{D}^{-1}    {\mathbold e}_i 
  \\
    &=     \frac {1-\rho} {\lambda_2 \, d_i \, h_i} ,
\end{align*}
where in the last step we used
$   {\mathbold e}_i'  
    \mathbold{D}^{-1}   \mathbold{A}  \mathbold{D}^{-1} \mathbold{A}    \mathbold{D}^{-1}    {\mathbold e}_i = (d_i h_i)^{-1}$.
Using our definitions
$
   \overline {\mathbold x}_i =  {\mathbold X}' \mathbold{B}
                 \mathbold{D}^{-1}{\mathbold e}_i 
$
and
$\mathbold{\Omega} = \mathbold{X}^\prime \mathbold{X}/m$
we obtain
\begin{align*}
       T^{(1)}_i &=  {\mathbold e}_i'   \mathbold{D}^{-1} 
                \mathbold{B}' {\mathbold X} \left(   {\mathbold X}' {\mathbold X} \right)^{-1}   {\mathbold X}' \mathbold{B}
                 \mathbold{D}^{-1}{\mathbold e}_i 
           = \frac 1 m \; \overline {\mathbold x}'_i  
         \, {\mathbold \Omega}^{-1}
             \overline {\mathbold x}_i      .
\end{align*}
Combining the above results we find
\begin{align*}
         {\rm var}\left(       \check\alpha_i \right)
            -  {\rm var}\left(       \hat \alpha_i \right)
         &\leq \frac{\sigma^2} \rho
           \left( T^{(1)}_i +  T^{(2)}_i + 2 \sqrt{T^{(1)}_i \, T^{(2)}_i}    \right)
      \\
       &\leq      \frac{\sigma^2} \rho
           \left(  \frac 1 m \; \overline {\mathbold x}'_i  
         \, {\mathbold \Omega}^{-1}
             \overline {\mathbold x}_i  +   \frac {1-\rho} {\lambda_2 \, d_i \, h_i} + 2 \sqrt{ \frac 1 m \; \overline {\mathbold x}'_i  
         \, {\mathbold \Omega}^{-1}
             \overline {\mathbold x}_i \,  \frac {1-\rho} {\lambda_2 \, d_i \, h_i}}    \right) .
\end{align*}
For any $a,b \geq 0$ we have
$a + b + 2 \sqrt{a b} \leq 2 (a+b)$. Thus, a slightly cruder but simpler bound is given by
     \begin{align*}
      \left|
            {\rm var}\left(       \check \alpha_i \right)
            -  {\rm var}\left(       \hat \alpha_i \right)
         \right|
        \; \leq  \;      \frac{2 \,\sigma^2} \rho
           \left(  \frac { \overline {\mathbold x}'_i   \, {\mathbold \Omega}^{-1}   \overline {\mathbold x}_i } m
             +   \frac {1-\rho} {\lambda_2 \, d_i \, h_i}   \right)  ,
     \end{align*}
where we also used that $ {\rm var}\left(       \check\alpha_i \right)
            \geq  {\rm var}\left(       \hat \alpha_i \right)  $, because adding regressors can only increase the variance
            of the least squares estimator under homoskedasticity.      \qed

\paragraph{PROOF OF THEOREM~\ref{thm:firstorderGeneralized}  (FIRST ORDER REPRESENTATION)}
\addcontentsline{toc}{subsection}{~~Proof of Theorem~\ref{thm:firstorderGeneralized}}
\mbox{}\newline
Remember that we treat $\mathbold{B}$ and $\mathbold{X}$ as fixed (i.e.\ non-random) throughout.
Let
$\check{\mathbold{\beta}} := \left( \mathbold{X}' \mathbold{M}_{\mathbold{B}}  \mathbold{X} \right)^{-1}  \mathbold{X}'  \mathbold{M}_{\mathbold{B}} \mathbold{y}  $. 
Using the model for $\mathbold{y} $ we find
$\check{\mathbold{\beta}}   -  {\mathbold{\beta}} = \left( \mathbold{X}' \mathbold{M}_{\mathbold{B}}  \mathbold{X} \right)^{-1}  \mathbold{X}'  \mathbold{M}_{\mathbold{B}} \mathbold{u}  $. 
Using our assumptions $ \mathbbmsl{E} (\mathbold{u}) = 0$
and $ \mathbold{\Sigma}  \leq   \mathbold{I}_m  \overline{\sigma}^2$
we find
$ \mathbbmsl{E}( \check{\mathbold{\beta}}  -  {\mathbold{\beta}} ) = 0$ and
\begin{align}
    \mathbbmsl{E} (\left( \check{\mathbold{\beta}}  -  {\mathbold{\beta}} \right) \left( \check{\mathbold{\beta}}  -  {\mathbold{\beta}} \right)')
    &= \left( \mathbold{X}' \mathbold{M}_{\mathbold{B}}  \mathbold{X} \right)^{-1}  \mathbold{X}'  \mathbold{M}_{\mathbold{B}}
         \mathbold{\Sigma} \mathbold{M}_{\mathbold{B}}  \mathbold{X}   \left( \mathbold{X}' \mathbold{M}_{\mathbold{B}}  \mathbold{X} \right)^{-1}
  \nonumber  \\
    &\leq   \overline{\sigma}^2   \left( \mathbold{X}' \mathbold{M}_{\mathbold{B}}  \mathbold{X} \right)^{-1}  \mathbold{X}'  \mathbold{M}_{\mathbold{B}}
         \mathbold{I}_m  \mathbold{M}_{\mathbold{B}}  \mathbold{X}   \left( \mathbold{X}' \mathbold{M}_{\mathbold{B}}  \mathbold{X} \right)^{-1}
    \nonumber \\
    &=       \overline{\sigma}^2   \left( \mathbold{X}' \mathbold{M}_{\mathbold{B}}  \mathbold{X} \right)^{-1} .
    \label{BoundBetaCheckVar}
\end{align}
The result in \eqref{PseudoinverseExpression} can be rewritten as
\begin{align}
    \mathbold{L}^\pinv
    =
     \left(  \mathbold{L}
    + m^{-1}  \mathbold{d}   \mathbold{d}^\prime  \right)^{-1}
    - m^{-1} \,  {\mathbold \iota}_n  {\mathbold \iota}_n'  .
  \label{PseudoinverseExpression2} 
\end{align}
The  constrained least-squares estimator in \eqref{eq:ols} can be expressed as
\begin{align} 
\check{\mathbold{\alpha}}
 =  
\; \;
\arg\hspace{-.9cm}
\min_{\mathbold{a}\in \lbrace \mathbold{a}\in\mathbb{R}^n: \,
 \mathbold{d}^\prime \mathbold{a}  = 0
 \rbrace}
\;\;
\left\lVert  \mathbold{y}-  \mathbold{X} \check{\mathbold{\beta}} -  \mathbold{B}\mathbold{a} \right\rVert^2,
    \label{DefAlphaCheckAlternative}
\end{align}
and analogous to Theorem~\ref{theorem:existence} we then find 
$
 \check{\mathbold{\alpha}}  
    = \mathbold{L}^\pinv \mathbold{B}' \left(  \mathbold{y}  - \mathbold{X} \check{\mathbold{\beta}} \right) 
  = \left(  \mathbold{L}
    + m^{-1}  \mathbold{d}   \mathbold{d}^\prime  \right)^{-1} \mathbold{B}' 
    \allowbreak \left(  \mathbold{y}  - \mathbold{X} \check{\mathbold{\beta}} \right) 
$.
Multiplying  by  $\left(  \mathbold{L}  + m^{-1}  \mathbold{d}   \mathbold{d}^\prime  \right)$ from the left and using
our normalization     $ \mathbold{d}^\prime   \check{\mathbold{\alpha}}  = 0$ gives
$$
\mathbold{L} \, \check{\mathbold{\alpha}}  = \mathbold{B}' \left(  \mathbold{y}  - \mathbold{X}  \check {\mathbold{\beta}} \right).
$$
Plugging $\mathbold{L}=\mathbold{D}-\mathbold{A}$ 
and $\mathbold{y} = \mathbold{B}  \mathbold{\alpha} +  \mathbold{X}  \mathbold{\beta} + \mathbold{u}$ 
into the last display, multiplying from the left with $\mathbold{D}^{-1}$, and rearranging terms,
we obtain
\begin{align}
    \check{\mathbold{\alpha}}  - \mathbold{\alpha} 
     =  \mathbold{D}^{-1}  \mathbold{B}' \mathbold{u} 
     +  \mathbold{\epsilon}  +  \tilde{\mathbold{\epsilon}}  ,
     \label{ExpansionCheckAlpha}
\end{align}
where
\begin{align*}
\mathbold{\epsilon} &:=     
\mathbold{D}^{-1}  \mathbold{A} \left( \check {\mathbold{\alpha}}  - \mathbold{\alpha} \right),
&
\tilde{\mathbold{\epsilon}} &:=  -   \mathbold{D}^{-1}  \mathbold{B}' \mathbold{X}  \left(  \check {\mathbold{\beta}} - {\mathbold{\beta}} \right) .
\end{align*}
We have $ \mathbbmsl{E} (\check{\mathbold{\beta}}  -  {\mathbold{\beta}} ) = 0$
and $ \mathbbmsl{E}  ( \check{\mathbold{\alpha}}  - \mathbold{\alpha} )  = 0$,
and, therefore, also
$\mathbbmsl{E} (\mathbold{\epsilon}) = \mathbf{0}$
and
$\mathbbmsl{E}( \tilde{\mathbold{\epsilon}} ) = \mathbf{0}$.
The definition $   \rho   = \left\lVert (\mathbold{X}^\prime \mathbold{X})^{-1}  \mathbold{X}^\prime \mathbold{M}_{\mathbold{B}} \mathbold{X} \right\rVert_2$
can equivalently be written as
$  \rho \mathbold{X}^\prime \mathbold{X} \geq   \mathbold{X}^\prime \mathbold{M}_{\mathbold{B}} \mathbold{X}   $,
and therefore $  \rho^{-1} \left( \mathbold{X}^\prime \mathbold{X} \right)^{-1} \leq 
 \left( \mathbold{X}^\prime \mathbold{M}_{\mathbold{B}} \mathbold{X} \right)^{-1}  $.
Using this and \eqref{BoundBetaCheckVar} we obtain
\begin{align*}
     \mathbbmsl{E}   (\tilde{\mathbold{\epsilon}}  \tilde{\mathbold{\epsilon}}^{\, \prime})
      &\leq   \overline{\sigma}^2  \mathbold{D}^{-1}  \mathbold{B}' \mathbold{X}
         \left( \mathbold{X}' \mathbold{M}_{\mathbold{B}}  \mathbold{X} \right)^{-1}
         \mathbold{X}'  \mathbold{B} \mathbold{D}^{-1}
     \\
     &\leq    \frac{  \overline{\sigma}^2 } \rho  \mathbold{D}^{-1}  \mathbold{B}' \mathbold{X}
         \left( \mathbold{X}'    \mathbold{X} \right)^{-1}
         \mathbold{X}'  \mathbold{B} \mathbold{D}^{-1} .
\end{align*}
Using
$
\check{\mathbold{\alpha}} -   \mathbold{\alpha} = (\mathbold{B}^\prime\mathbold{M}_{\mathbold{X}}\mathbold{B})^{\pinv} \mathbold{B}^\prime \mathbold{M}_{\mathbold{X}}\mathbold{u} 
$
and the assumption  $ \mathbold{\Sigma} \leq \overline{\sigma}^2 \mathbold{I}_n$ we calculate
\begin{align*}
     \mathbbmsl{E}  (\mathbold{\epsilon}  \mathbold{\epsilon}' )
     &=    \mathbold{D}^{-1}  \mathbold{A}   (\mathbold{B}' \mathbold{M}_{\mathbold{X}} \mathbold{B})^\pinv \mathbold{B}' \mathbold{M}_{\mathbold{X}}
     \mathbold{\Sigma}
        \mathbold{M}_{\mathbold{X}}  \mathbold{B} (\mathbold{B}' \mathbold{M}_{\mathbold{X}} \mathbold{B})^\pinv     \mathbold{A} \mathbold{D}^{-1} 
     \\
     &\leq     \overline \sigma^2  \mathbold{D}^{-1}  \mathbold{A}   (\mathbold{B}'  \mathbold{M}_{\mathbold{X}} \mathbold{B})^\pinv \mathbold{B}'  
          \mathbold{M}_{\mathbold{X}}  \mathbold{B} (\mathbold{B}'  \mathbold{M}_{\mathbold{X}} \mathbold{B})^\pinv     \mathbold{A} \mathbold{D}^{-1} 
     \\
     &=       \overline \sigma^2 \, \mathbold{D}^{-1}  \mathbold{A}     (\mathbold{B}'  \mathbold{M}_{\mathbold{X}} \mathbold{B})^\pinv  \mathbold{A} \mathbold{D}^{-1} 
   \\
    &\leq 
       \overline \sigma^2 \, \mathbold{D}^{-1}  \mathbold{A}     (\mathbold{B}'    \mathbold{B})^\pinv  \mathbold{A} \mathbold{D}^{-1}  
       +  \frac{ \overline \sigma^2} {\rho} \, \mathbold{D}^{-1}  \mathbold{A} 
            \left(\mathbold{B}'  \mathbold{B}  \right)^\pinv
\mathbold{B}' {\mathbold X}    \left(   {\mathbold X}' {\mathbold X} \right)^{-1}   {\mathbold X}'  \mathbold{B}
  \left(\mathbold{B}'  \mathbold{B}    \right)^\pinv
            \mathbold{A} \mathbold{D}^{-1} ,
\end{align*}
where in the last step we used \eqref{BoundBMB}.
Since furthermore
 ${\mathbold X}    \left(   {\mathbold X}' {\mathbold X} \right)^{-1}   {\mathbold X}'  \leq {\mathbold I}_m$
 and  $ \left(\mathbold{B}'  \mathbold{B}  \right)^\pinv = \mathbold{L}^\pinv \leq \lambda_2^{-1} \mathbold{D}^{-1}$
we obtain
\begin{align*}
     \mathbbmsl{E} ( \mathbold{\epsilon}  \mathbold{\epsilon}' )
     &\leq 
      \overline \sigma^2 \, \mathbold{D}^{-1}  \mathbold{A}     (\mathbold{B}'    \mathbold{B})^\pinv  \mathbold{A} \mathbold{D}^{-1}  
       +  \frac{ \overline \sigma^2} {\rho} \, \mathbold{D}^{-1}  \mathbold{A} 
            \left(\mathbold{B}'  \mathbold{B}  \right)^\pinv
\mathbold{B}'   \mathbold{B}
  \left(\mathbold{B}'  \mathbold{B}    \right)^\pinv
            \mathbold{A} \mathbold{D}^{-1}
     \\       
     &=      \frac{ \overline \sigma^2 (1+\rho)} \rho \, \mathbold{D}^{-1}  \mathbold{A}     (\mathbold{B}'    \mathbold{B})^\pinv  \mathbold{A} \mathbold{D}^{-1}  
     \\
      &\leq    \frac{ \overline \sigma^2 (1+\rho)} {\lambda_2 \, \rho} \, \mathbold{D}^{-1}  \mathbold{A}    \mathbold{D}^{-1}\mathbold{A} \mathbold{D}^{-1}  .
\end{align*}
Denote the elements of  $\mathbold{\epsilon}$ and $\tilde{\mathbold{\epsilon}}$ by $ \epsilon_i $
and  $\tilde \epsilon_i $.
Equation \eqref{ExpansionCheckAlpha} can then be written as
\begin{align*}
 \check{\alpha}_i-\alpha_i &= \frac{\mathbold{b}^\prime_i \mathbold{u}}{d_i} 
 + \epsilon_i + \tilde \epsilon_i     ,
\end{align*}
and we have
\begin{align*}
     \mathbbmsl{E} (\epsilon_i^2 )\leq
       \frac{ \overline \sigma^2 (1+\rho)} {\lambda_2 \, \rho} \,
     \mathbold{e}_i'   \mathbold{D}^{-1}  \mathbold{A}    \mathbold{D}^{-1}\mathbold{A} \mathbold{D}^{-1}   \mathbold{e}_i
     =    \frac{ \overline \sigma^2 (1+\rho)} {\lambda_2 \, \rho} \,
      \frac{1} {d_i \, h_i} ,
\end{align*}
and
\begin{align*}
     \mathbbmsl{E} ( \tilde \epsilon_i^{\,2} )
     &\leq
      \frac{  \overline{\sigma}^2 } \rho 
         \mathbold{e}_i'    \mathbold{D}^{-1}  \mathbold{B}' \mathbold{X}
         \left( \mathbold{X}'    \mathbold{X} \right)^{-1}
         \mathbold{X}'  \mathbold{B} \mathbold{D}^{-1}   \mathbold{e}_i
         =  \frac 1 m \;  \frac{  \overline{\sigma}^2 } \rho \; \overline{\mathbold{x}}_i'  \, \mathbold{\Omega}^{-1} \, \overline{\mathbold{x}}_i .
\end{align*}     
where we used our
definitions
$\overline{\mathbold{x}}_i   =   {\mathbold X}' \mathbold{b}_i / d_i
=   \mathbold{X}'  \mathbold{B} \mathbold{D}^{-1}   \mathbold{e}_i$
 and $\mathbold{\Omega}:= \mathbold{X}^\prime \mathbold{X}/m$.  \qed

\paragraph{PROOF OF THEOREM~\ref{thm:inid} (ASYMPTOTIC DISTRIBUTION)}
\addcontentsline{toc}{subsection}{~~Proof of Theorem~\ref{thm:inid}}
\mbox{}\newline
We have $\rho \leq 1$ by definition.
Together with the assumptions $\overline{\sigma}^2 = O(1)$, $\lambda_2 h_i\rightarrow\infty$,
and the conditions in \eqref{AsymptoticConditionsBeta} 
this implies that
$\mathbbmsl{E}(\epsilon_i^2)\leq \overline{\sigma}^2 (1+\rho) /(\rho \, d_i \, \lambda_2 \, h_i) = o(d_i^{-1})$,
and $\mathbbmsl{E}(\tilde \epsilon_i^{\,2})\leq 
     \overline{\sigma}^2  \,   \overline{\mathbold{x}}_i'  \, \mathbold{\Omega}^{-1} \, \overline{\mathbold{x}}_i  / (\rho \, m) = o(d_i^{-1})$.
By  Markov's inequality we thus have $\epsilon_i = o_p(d_i^{-1/2})$ and $\tilde \epsilon_i = o_p(d_i^{-1/2})$,
and applying Theorem~\ref{thm:firstorderGeneralized} gives, as  $d_i\rightarrow\infty$,
\begin{align*}
(\check{\alpha}_i-\alpha_i )
&\overset{p}{\rightarrow}
\frac{ \mathbold{b}^\prime_i \mathbold{u}   } {d_i} = \frac{1}{ d_{i}} \sum_{j \in [i]} \sum_{e \in E_{(i,j)}} \nu_{\varepsilon_{e} i} ,
&
\nu_{\varepsilon_{e} i}
&:=
 \left( \mathbold{B} \right)_{\varepsilon_{e} i}  u_{\varepsilon_{e}} .
\end{align*}
The number of terms $\nu_{\varepsilon_{e} i} $ summed over in the last display grows to infinity asymptotically, because
we assume that $d_i =  \sum_{j \in [i]} \sum_{e \in E_{(i,j)}}  w_e \rightarrow \infty$,
while the weights $w_e = \left( \mathbold{B} \right)_{\varepsilon_{e} i}^2$ are bounded.
Our assumptions furthermore guarantee that the $\nu_{\varepsilon_{e} i}$ are independent and satisfy
$\mathbbmsl{E}  (\nu_{\varepsilon_{e} i}) = 0$,
 $\mathbbmsl{E}  (\nu_{\varepsilon_{e} i}^2 )\geq c_1 > 0$, and
$\mathbbmsl{E} (\left| \nu_{\varepsilon_{e} i} \right|^3) \leq c_2 < \infty$
for constants $c_1,c_2$. Thus, the Lyapunov condition is satisfied, and 
the statement of the theorem then follows from a standard application of Lyapunov's central limit theorem.
\qed

\end{document}